\documentclass[twocolumn,aps,pre,amssymb,showpacs]{revtex4}
\usepackage{graphicx}
\usepackage{graphics}
\usepackage{epsfig}

\begin{document}

\title{Extraction of coherent structures in a rotating turbulent flow experiment}
\author{Jori E. Ruppert-Felsot, Olivier Praud, Eran Sharon\footnote{current affiliation: The Racah Institute of Jerusalem,
Jerusalem, Israel}, Harry L.
Swinney\footnote{swinney@chaos.ph.utexas.edu}}

\affiliation{Center for Nonlinear Dynamics and Department of
Physics, The University of Texas at Austin, \\ Austin, Texas  78712}



\begin{abstract}

The discrete wavelet transform (DWT) and discrete wavelet packet
transform (DWPT) are used to extract and study the dynamics of
coherent structures in a turbulent rotating fluid. Three-dimensional
(3D) turbulence is generated by strong pumping through tubes at the
bottom of a rotating tank (48.4 cm high, 39.4 cm diameter). This
flow evolves toward two-dimensional (2D) turbulence with increasing
height in the tank.  Particle Image Velocimetry (PIV) measurements
on the quasi-2D flow reveal many long-lived coherent vortices with a
wide range of sizes. The vorticity field exhibits vortex birth,
merger, scattering, and destruction. We separate the flow into a
low-entropy ``coherent'' and a high-entropy ``incoherent'' component
by thresholding the coefficients of the DWT and DWPT of the
vorticity field. Similar thresholdings using the Fourier transform
and JPEG compression together with the Okubo-Weiss criterion are
also tested for comparison. We find that the DWT and DWPT yield
similar results and are much more efficient at representing the
total flow than a Fourier-based method. Only about 3\% of the
large-amplitude coefficients of the DWT and DWPT are necessary to
represent the coherent component and preserve the vorticity
probability distribution function (PDF), transport properties, and
spatial and temporal correlations. The remaining small amplitude
coefficients represent the incoherent component, which has near
Gaussian vorticity PDF, contains no coherent structures, rapidly
loses correlation in time, and does not contribute significantly to
the transport properties of the flow. This suggests that one can
describe and simulate such turbulent flow using a relatively small
number of wavelet or wavelet packet modes.
\end{abstract}


\maketitle

\section{Introduction}

Large-scale ordered coherent motions occur in a wide variety of
turbulent flows despite existing in a rapidly fluctuating background
turbulence~\cite{hussain.86,sreenivasan.99}. These ``coherent
structures'', which are associated with localized regions of
concentrated vorticity, persist for times that are long compared to
an eddy turnover time. The importance of coherent structures in
turbulence has become recognized through the use of flow
visualization. Examples of coherent structures identified in
turbulent flows include hairpin vortices in boundary layer
turbulence~\cite{robinson.91}, plumes in turbulent convection, and
vortices in turbulent shear flows. Coherent structures play a major
role in the transport of mass and momentum, thus affecting
transport, drag, and dissipation in turbulent flows. Due to their
long lifetimes, coherent structures in the atmosphere strongly
influence the exchange of heat, moisture and nutrients between
different locations.  In industry, the prediction and control of
transport, drag, and turbulence is important in many
processes~\cite{hunt.91}, and in some cases flows can be modified
through the control of coherent structures~\cite{liu.89,lumley.98}.

The emergence of coherent vortices is especially striking in
geostrophic turbulence, where the Coriolis force plays a dominant
role~\cite{sommeria.88,mcwilliams.84}. Under the effect of rotation
and stratification, geophysical flows develop large and robust
coherent structures which can be identified and tracked for time
much longer than their characteristic turnover time. Examples of
such flows are high and low pressure systems, large vortical
structures that are formed in Gulf Stream meander, Mediterranean
eddies (Meddies), and the Earth's jet stream; all can be observed
and tracked by satellite imaging of the atmosphere or the ocean
surface (e.g. see satellite imaging
websites~\cite{satellite.1,satellite.2}). Large coherent structures
are not limited to the Earth; the atmospheres of other planets also
reveal structures such as Neptune's dark spot and Jupiter's zones,
belts and Great Red Spot.

Three-dimensional (3D) turbulence subjected to strong rotation
develops columnar vortical structures aligned with the rotation
axis, as observed in the present experiment and in previous
laboratory
experiments~\cite{mcewan.76,verdiere.80,hopfinger.82,dickinson.83}.
Correlation in this direction becomes large when the Coriolis force
becomes large compared to inertial forces, and the flow proceeds
towards a quasi-2D state. This two-dimensionalization allows energy
to proceed toward larger scales through the inverse energy cascade
and toward smaller scales through the forward enstrophy
cascade~\cite{kraichnan.67}, as observed in
simulation~\cite{smith.96,smith.99,godeferd.99} and experiments on
rotating~\cite{verdiere.80,hopfinger.82} and non-rotating quasi-2D
flows~\cite{tabeling.97}. The cascades of energy and enstrophy lead
to a spontaneous appearance of intense localized coherent vortices
containing most of the enstrophy of the flow. Vorticity filaments
outside of those coherent structures are distorted and advected by
the velocity field induced by the vortices. Thus, a rotating
turbulent fluid flow can organize itself into large-scale coherent
structures that are often long-lived compared to dissipative time
scales. The presence of long-lived coherent structures in turbulent
rotating flow has been observed in both
experiment~\cite{mcewan.76,verdiere.80,hopfinger.82,sommeria.88} and
in simulation~\cite{smith.99,godeferd.99}. These structures are
larger than the scale of the forcing. The large-scale coherent
structures break the homogeneity of the flow and are thought to
dominate the flow dynamics. The long lifetimes and spatial extent
allow coherent structures to play a significant active role in
transport processes \cite{provenzale.99,solomonweeks.93}.

Due to the dynamical importance of coherent structures, any
analysis of flow containing these structures should take into
account their existence. One approach to analyze the coherent
structures and the dynamics of such flow is to partition it into
regions with different dynamical properties. Okubo \cite{okubo.70}
derived a criterion to separate flow into a region where strain
dominates (hyperbolic region) and a region where vorticity
dominates (elliptic region). The same criterion was later
re-derived by Weiss \cite{weiss.91} and is now known as the
Okubo-Weiss criterion. The criterion has been widely used to
analyze numerical simulations of 2D turbulence. However, as
pointed out by Basdevant and Philipovitch \cite{basdevant.94}, the
validity of the criterion's key assumption is restricted to the
core of the vortices that correspond to the strongest elliptic
regions. This limitation reduces the applicability of a
decomposition using this criterion.

Another method that has been found useful in analyzing flow fields
is proper orthogonal decomposition (POD). This projects a field onto
a set of orthonormal basis functions where successive eigenvectors
are obtained by numerically maximizing the amount of energy
corresponding to that eigenmode~\cite{holmeslumley}. [POD is known
by other names, including Karhunen-Lo$\rm\grave{e}$ve decomposition,
principal components analysis (PCA), and singular value
decomposition (SVD); the basis functions are also known as empirical
eigenfunctions and empirical orthogonal
functions~\cite{holmeslumley}, p.86.] Linear combinations of the
basis functions ideally correspond to the coherent structures in the
flow. However, if the ensemble of fields is homogeneous, the basis
functions become Fourier modes~\cite{holmeslumley}. Furthermore, a
single mode of the POD corresponds to a full field with structures
in a particular spatial arrangement. If the structures in the field
are not stationary, as in the case of our present flow, many modes
will therefore be necessary to track the different spatial
configurations. Indeed, application of POD to our data resulted in
the extraction of the large scale, low amplitude mean flow structure
of our field, rather than the continually moving intense coherent
structures. We therefore do not include POD in this paper.

This paper separates a flow into coherent and incoherent components
using wavelet transforms~\cite{daubechies} to extract localized
features at different spatial scales. The most important advantage
of the wavelet representation over the more usual Fourier
representation is the localization of the basis functions. A Fourier
analysis is not well suited to pick out localized features such as
intense vortices. The basis functions of a Fourier transform are
localized in wavenumber space and hence spread out over the entire
domain in physical space. The basis functions of the wavelet
transform consist of dilates and translates of a ``mother'' wavelet,
which contains multiple frequencies and has compact support
(non-zero values only inside a finite interval) in physical space.
The basis functions are well localized in both physical and
wavenumber space; hence only a few coefficients suffice to describe
localized features of a signal. The coefficients of the wavelet
transform contain not only amplitude but also scale and position of
the basis elements. Thus the coefficients can be used to track the
size and location of features that are well correlated with the
wavelet bases (e.g. \cite{weiss.97,schram.2001}).

The discrete wavelet transform (DWT) \cite{daubechies} has been
found to be well suited to analyze intermittent signals and systems
containing localized features such as the intense vortices that
occur in turbulence \cite{farge.92rev,farge.92,farge.99}. Farge has
extensively applied wavelets to the analysis and computation of
turbulent
flows~\cite{farge_website,farge.92rev,farge.92,farge.99,farge.2003,beta.2003}.

The discrete wavelet \textit{packet} transform
(DWPT)~\cite{wickerhauser} is a generalization of the DWT; the
possible wavelet packet basis elements are a larger set which
include spatial modulation of the mother wavelet. The advantage of
the DWPT is that the choice of basis is adaptable to the signal to
be analyzed.

Turbulent flow can be considered as a superposition of large-scale
coherent motions, ``fine-scale" incoherent turbulence, and a mean
flow with interaction between the three
constituents~\cite{hussain.83}. In numerical simulations of 2D
turbulence~\cite{farge.92,weiss.97,farge.99}, and more recently 3D
turbulence~\cite{farge.2003}, the coherent and incoherent turbulent
background components have been separated using wavelet based
decompositions operating on the vorticity field. The coherent part,
represented by only a small fraction of the coefficients, retained
the total flow dynamics and statistical properties, while the
incoherent part represented no significant contribution to the flow
properties. This separation of the flow into two dynamically
different components suggests that the computational complexity of
turbulent flows could be reduced in simulations with coherent
structures interacting with a statistically modelled incoherent
background~\cite{farge.99}. The application, however, has been
heretofore primarily limited to results obtained from numerical
simulations.

In this paper, we use the wavelet technique to analyze our
Particle Image Velocimetry (PIV) data on a rapidly rotating
turbulent flow. Section~\ref{experiment} describes the
experimental system. Section~\ref{flowfields} describes the
resulting flow fields obtained in the experiment.
Section~\ref{decomposition} presents the techniques used to
decompose the vorticity field into coherent and incoherent
components. Section~\ref{results} presents the results obtained by
applying the method to measurements on rotating turbulent flow.
The conclusions are discussed in Section~\ref{discussion}.

\section{Experiment}
\label{experiment}

\subsection{Instrumentation and measurements}

\begin{figure}[h!]
\begin{center}
  \includegraphics[width=80mm]{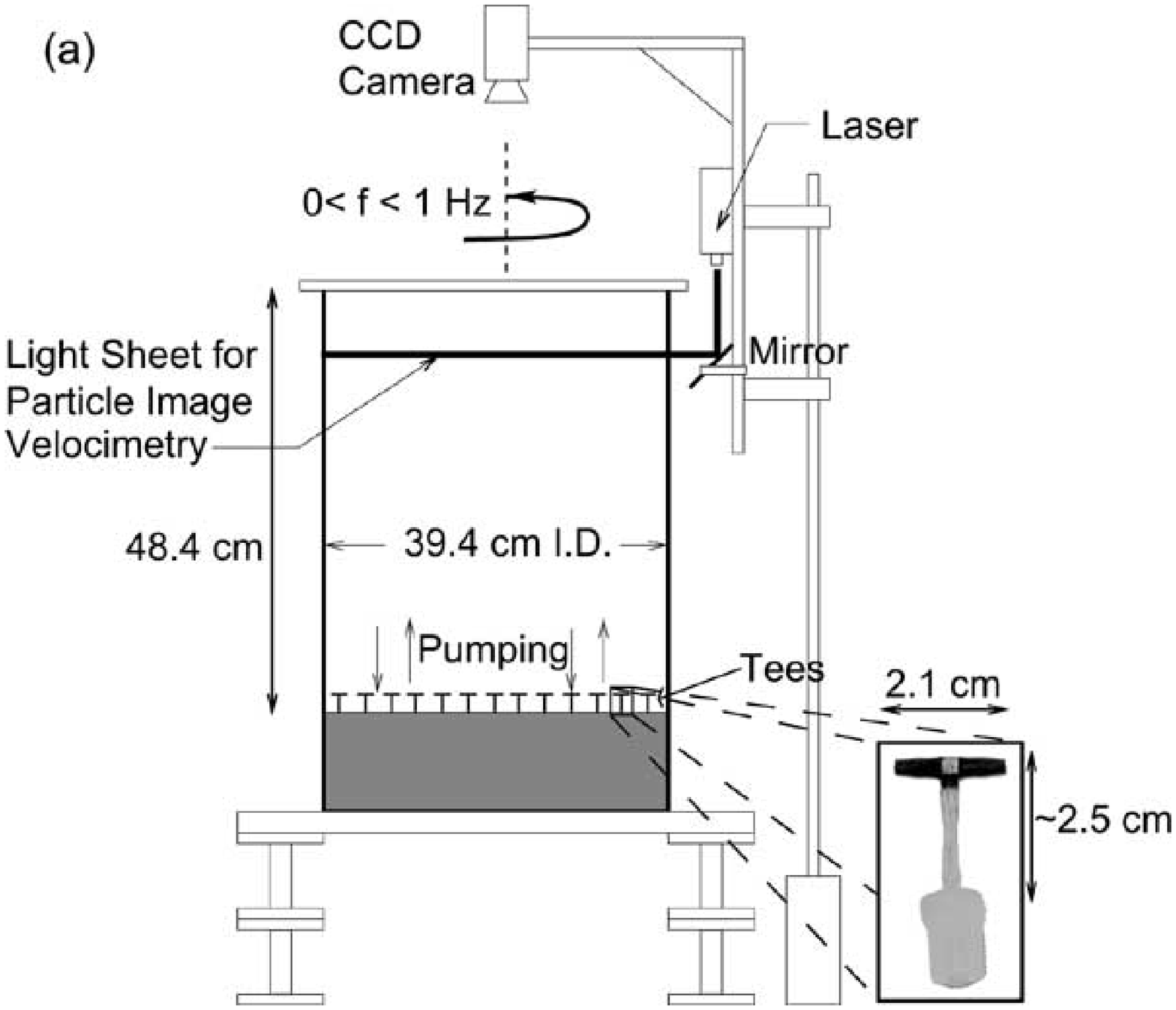}\\
\end{center}

  \includegraphics[width=80mm]{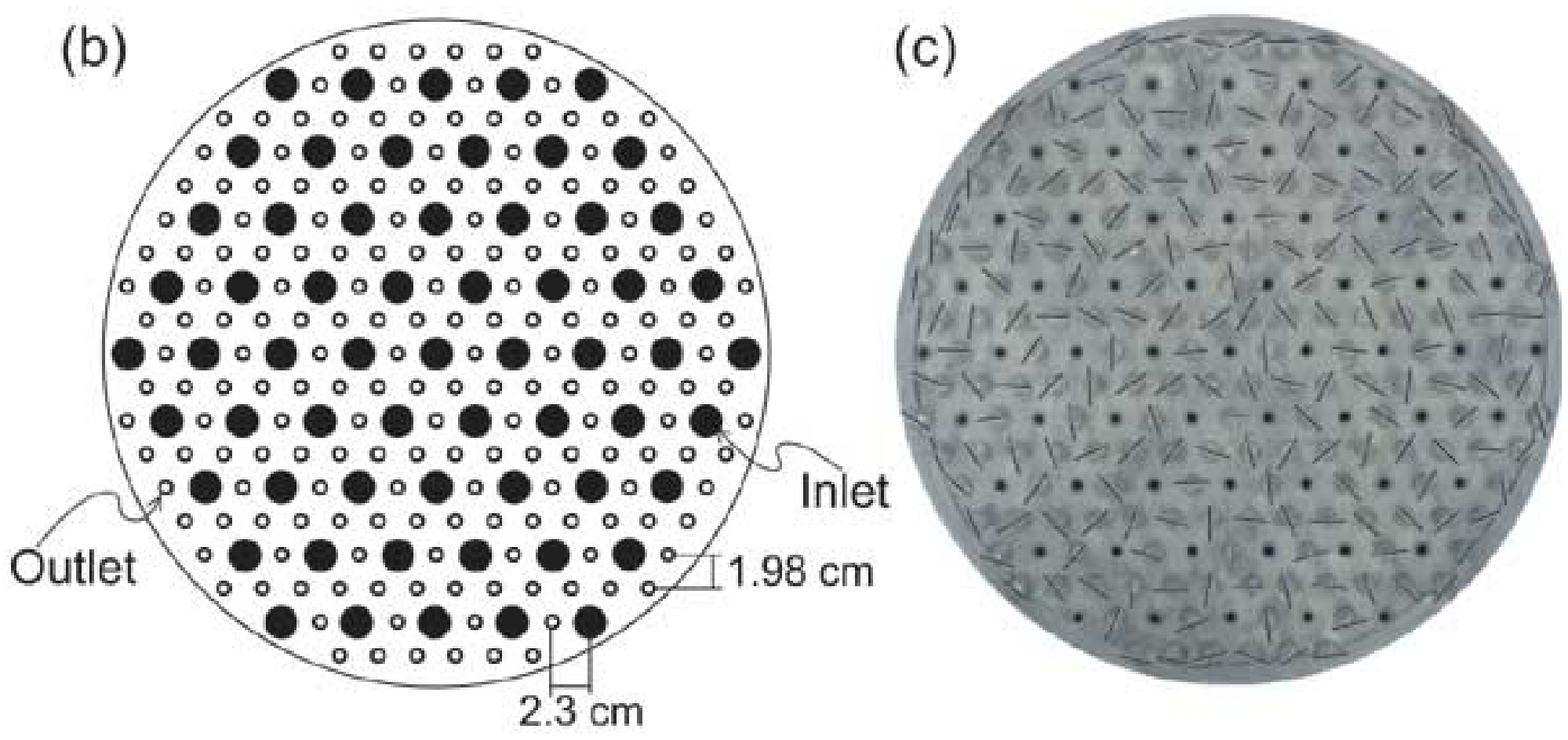}
  \caption{ (a) A schematic of the apparatus. The height of the laser
sheet is adjustable. Shown in the inset is a close-up a single
tee. (b) Horizontal cross section of the distribution of pumping
  sources (open circles) and sinks (black dots) (positions are to
  scale; see text for relative size of sources and sinks and tank diameter scale).
  (c) Overhead image of tank bottom, showing the tees. Overlayed are lines
  indicating the orientation of the tees.}
\label{apparatus}
\end{figure}

An acrylic cylinder (48.4 cm tall, 39.4 cm inner diameter) is fit
inside a square acrylic tank ($40\times40$ cm cross section, 60 cm
tall) that has a transparent lid
(Fig.~\ref{apparatus})~\cite{mastersthesis}. The tank is filled
with distilled water at $24\pm1 ^\circ$C ($\rho$ = 0.998 g/cm$^3$,
$\nu$ = 9.5 $\times 10^{-3}$ cm$^2$/s). The tank and the data
acquisition computer are mounted on a table that can be rotated up
to 1.0 Hz.

Fluid is injected at the bottom of the tank by pumping water through
a hexagonal array of tubes with grid spacing 2.3 cm; there are 192
sources and 61 sinks, as shown in Fig.~\ref{apparatus}. The source
tubes are tee-shaped with a 2.5 cm shaft and 2.1 cm horizontal top
(0.079 cm inside diameter) [see inset of Fig.~\ref{apparatus}(a)].
Each tee is screwed into the distributor. This design allows us to
easily change the type of connector or change the forcing geometry
by blocking or modifying sources or sinks. The tee source geometry
was chosen to produce horizontal velocities and to minimize the
vertical motions directly driven by the forcing. Further, the
orientation of the tees was chosen to minimize cooperation between
neighboring tees; however, it was not possible to eliminate
completely a mean flow. The sinks are 1.27 cm diameter holes in the
bottom of the distributor [Fig.~\ref{apparatus}(b)]. The forcing
system is versatile and allows us to inject nearly homogeneous
turbulence at a scale much smaller than the system size.

Flow rates can range up to 1400 cm$^3$/s, corresponding to flow
velocities up to 8 m/s out of the tees. This corresponds to a
Reynolds number of order 10$^5$ based on the jet velocity and the
grid spacing of the tees. The Rossby number, U/2$\Omega$L, based
upon the same scales is order 50 near the bottom of the tank, near
the forcing.  However, the turbulence decays away from the
forcing, with increasing height, and becomes more 2D due to the
influence of rotation.

The data presented are for 0.4 Hz rotation rate and flow rates of
426 cm$^3$/s, producing 2.4 m/s velocity jets at the forcing tees.
This corresponds to Reynolds and Rossby numbers of about
$6\times10^4$ and $20$ respectively, near the forcing. Near the
top the RMS flow velocity is about 2 cm/s and the maximal
characteristic velocity length scale (twice the $e$-folding length
of the velocity spatial correlation function) is about 5 cm; hence
near the top the Reynolds number is about 1000. The Rossby number
calculated from characteristic scales is 0.08, while the local
averaged Rossby number calculated from the RMS vorticity,
$(\omega_{RMS}/2\Omega)$, at such a flow condition is 0.3. The
depth of our system thus allows us to observe more 3D (Rossby
$\approx$ 20) or more 2D (Rossby $\approx$ 0.3) flow without
changing control parameters, such as rotation or forcing.

The water is seeded with polystyrene ($\rho$ = 1.067 g/cm$^3$)
spherical particles with diameters in the range 90-106 $\mu$m. The
small mismatch in density results in a sedimentation terminal
velocity of $4\times10^{-2}$ cm/s in the absence of flow, which is
insignificant compared to the measured flow velocities. The
density mismatch also causes a lag in the response of the
particles in regions of large acceleration
\cite{piv,michaelides.97}. We estimate the largest lag to be a 1\%
difference between the particle velocity and the flow velocity in
the vortices at the top of the tank.

A 395 mW, 673 nm diode laser with attached light-sheet optics
(from Lasiris) illuminates the particles in horizontal planes for
flow visualization and PIV measurements. It is also possible to
rotate the laser and optics by 90$^\circ$ to illuminate vertical
planes for measurements of the vertical velocities. The thickness
of the sheet is about 1 cm and varies less than 10\% across the
diameter of the tank. The laser is fixed to a carriage which
allows us to adjust the vertical position of the light-sheet
without changing the pumping rate or system rotation rate.

Particles are imaged in the rotating frame with a CCD camera
($1004\times1004$ pixels, 30 frames per second) mounted above the
tank. The images are grabbed and stored into a memory buffer by a
computer on the rotating table.  A second, lower resolution camera
with an analog output allows us to view the flow when not grabbing
digital images.

The laser pulses were timed so that pairs of pulses were imaged in
successive frames of the CCD camera, which had a dead time of 120
$\mu$s between consecutive frames. The pulse duration was
typically 10 ms with a 2 ms dark interval between pulses.

The analog signal from the second camera and power for the equipment
on the table are sent through slip rings. The entire experiment is
controlled remotely with a wireless ethernet connection and remote
access software (PC Anywhere) by a computer in the non-rotating
laboratory frame. This allows us to take multiple data sets, adjust
the pumping rate, and adjust the position of the light sheet without
having to stop the table.

To determine the two-dimensional projection of the velocity field
we use a type of Digital Particle Image Velocimetry (DPIV) known
as Correlation Image Velocimetry
(CIV)~\cite{fincham.97,fincham.2000}.  The CIV algorithm uses
cross-correlations over small interrogation regions between a pair
of consecutive images to find particle displacements over known
time intervals to obtain velocities. The CIV algorithm also allows
for possible rotations and deformations of the interrogation
region to take into account nonlinear particle motions in between
images. The resulting velocity field is interpolated to a uniform
regular grid by a cubic spline interpolation. The vorticity fields
are calculated from the coefficients of the spline fit.

We can achieve a maximum data rate of 15 velocity fields per second
(using two images to calculate each velocity field). This time
resolution is sufficient for our flow conditions. To capture the
dynamics of the flow, data were taken for 20 s periods (limited by
memory buffer size), resulting in 300 fields in a single sequence.
To capture statistical information, data were taken at 30 s
intervals (longer than measured decay time for our flow) over (in
most cases) runs of 50 minutes in duration. This was sufficient to
achieve convergence of statistical quantities. The resulting
velocity and vorticity fields are on a 128$\times$128 grid with a
spatial resolution of 0.30 cm when the whole tank is imaged and 0.19
cm when we zoom-in. Testing the CIV algorithm against test particle
images in known flow fields
\cite{okamoto.2000,PIV_standard_website}, we found that the velocity
fields determined by the algorithm have about 2\% RMS uncertainty.

\subsection{Decay time}
\label{decaytime}

Characteristic decay times for our flow were measured by three
different types of experiments. In each experiment, the measured
decay time was taken as the $e$-folding relaxation time of the
mean kinetic energy in the flow.

(i) Laminar (no forcing) \textit{spin-down} experiments were
conducted with the tees removed and replaced with a flat
horizontal boundary. The tank was subjected to a sudden 10\%
decrease in rotation rate after the flow had reached solid-body
rotation (no motion in the rotating tank frame). The predicted
decay time of motions in laminar rotating flows with rigid, flat
horizontal boundaries subject to small step changes in rotation is
given by the \textit{Ekman} dissipation time, $\tau = H/(2
\sqrt{\nu \Omega})$, where $H$ is the depth of the fluid, $\nu$ is
the kinematic viscosity and $\Omega = 2\pi f$ is the angular
frequency of the container~\cite{pedlosky}. The corresponding
decay time for the energy in the flow is $\tau = H/(4 \sqrt{\nu
\Omega})$. For our closed cylindrical tank without topography of
depth H = 48.4 cm, $\nu = 0.095$ cm$^2$/s, and a rotation rate of
0.4 Hz, this gives a decay time of the energy as 78 s; the
measured time was 64 s.

(ii) Laminar spin-down experiments were conducted in the same way
as in (i), except that the tees were installed in the bottom of
the tank. The characteristic decay time measured with the tees
installed in the bottom of the tank for our flow was 18$\pm$3 s.
The reduced laminar spin down time is due to the tees, which cause
extra drag and secondary circulations that quickly bring the fluid
to solid-body rotation.

(iii) Turbulent decay experiments were conducted by abruptly
shutting off the forcing. For these experiments, the flow was
allowed to reach a steady turbulent state (typically many decay
times) under a constant pumping rate of 426 cm$^3$/s and rotation
rate of 0.4 Hz before abruptly turning off the forcing. The
measured decay time of 13$\pm$3 s is long compared to the typical
vortex turnover time of 1 s.

\subsection{Passive scalar advection}

To study the transport and mixing properties of our flow we
examine numerically the motion of passive scalar point particles
and passive scalar fields in the velocity fields obtained from the
experiment. The velocity fields that we measure are a
two-dimensional projection of a three-dimensional incompressible
flow field. Therefore, they have a non-zero divergence and do not
satisfy any fluid dynamical equation of motion. Nonetheless, they
can give us useful information regarding the transport properties
of the flow.

Initial locations are chosen for the point particles, the
positions of the particles are updated corresponding to the
velocity fields such that $\mathbf{x}_{n+1} = \mathbf{u}_n \Delta
t + \mathbf{x}_n$ where $\mathbf{x}_n$ is the position of the
particle at time-step $n$, and $\mathbf{u}_n$ is the velocity of
the flow field at time-step $n$ at the location of the particle.
Each velocity field is interpolated in space by a cubic-spline to
calculate the field at the location of a given particle. The
timestep, $\Delta t$, is chosen by the Courant condition
\cite{numericalrecipes}, which avoids particles jumping over grid
points or going too far at a given iteration. For our data this
condition means $\Delta t < 0.02$ s $<$ 1/15 s; therefore, fields
must also be interpolated in time. The measured velocity fields
vary slowly in time compared to our temporal resolution 1/15 s
(measured correlation time is $\approx 2.4$ s; see
section~\ref{decmpvortfields}). We therefore justify the use of a
cubic spline interpolation in time to achieve $\Delta t$ $\ll$
1/15 s (typically $\Delta t = 0.001$ s).

We also examine the time evolution of a passive scalar field
advected in the experimental velocity field by numerically
integrating the advection-diffusion equation,

\begin{equation}\label{advecdiff}
\frac{\partial c}{\partial t} = - \mathbf{u}\cdot\nabla c +
\kappa_D \nabla^2 c,\end{equation}

\noindent where $c = c(\mathbf{x},t)$ is the passive scalar
concentration and $\kappa_D$ is a diffusion coefficient, chosen as
necessary for numerical stability of the solution. The values of
$\kappa_D$ used correspond to Schmidt numbers ($\kappa_D$/$\nu$)
near 0.05. The numerical integration is performed using a
pseudo-spectral method in polar coordinates based upon methods
given in \cite{trefethen}. The grid is 128 Chebyshev modes in the
radial direction and 256 Fourier modes in the azimuthal direction.
The velocity field obtained by the CIV measurement is interpolated
via a cubic spline onto the simulation grid. The numbers of radial
and azimuthal modes were chosen so as to not under-resolve our
velocity fields on the nonuniform polar grid.

The singularity at the origin was avoided by choosing Chebyshev
modes for the radial direction. The collocation points for Chebyshev
modes have unit spacing on a circle, $x_i = \cos{(\pi i/(N-1))}$,
where $x_i$ is the $i^{\mathrm{th}}$ gridpoint, and $N$ the total
number of gridpoints. The gridpoints are clustered at the boundaries
and sparse in the center of the domain. By defining the radial
coordinate $r \in [-1,1]$ and using only the $r > 0$ values, the
radial grid points are sparse near $r = 0$ and clustered at the
boundary $r = R$. For an even number of modes the origin, $r = 0$,
is skipped.  However, the azimuthal grid is still dense at the
origin, which requires us to use a very short time step $\Delta t
\approx 2 \times 10^{-4}$ s to avoid numerical instability. The
experimentally determined velocity fields are then interpolated in
time by the method described above for the tracer particle
simulation.

The diffusive term is calculated implicitly by a Crank-Nicholson
scheme, separately for the radial and azimuthal directions. The
advection term is calculated explicitly by a predictor-corrector
scheme using a third-order Adams-Bashforth step followed by a
fourth-order Adams-Moulton step \cite{numericalrecipes}.

To perform the numerical analysis on the decomposed fields we
construct a velocity field from a vorticity field. We make the
assumption of 2D flow that the total vorticity is given by the
measured vertical vorticity, $\omega = \omega_z$. We then use the 2D
streamfunction-vorticity relation $\nabla^2 \Psi = -\omega_z$ and
solve Poisson's equation for the streamfunction $\Psi$ using
Matlab's PDE solver. The derivatives of the streamfunction are then
used to calculate the velocity field by the relation $\nabla \times
(\Psi \mathbf{z}) = \mathbf{u}$, where $\mathbf{z}$ is the unit
vector in the vertical direction.

\begin{figure}
  \includegraphics[width=83mm]{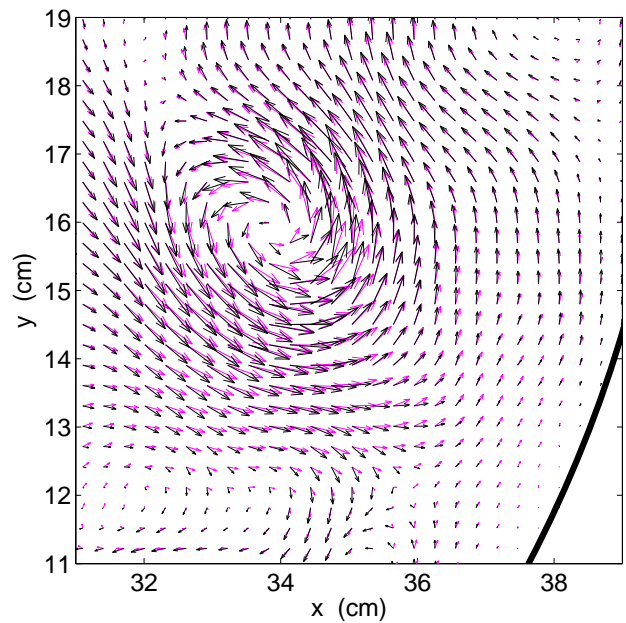}\\
  \caption{(color on-line) Close-up of a vortex in our tank near the boundary (bold line).
  The original field is represented by black vectors and
  the reconstructed field by light-colored vectors. The original
  and reconstructed fields are indistinguishable in the regions of uniform flow.
  The largest vectors correspond to 6.7 cm/s.
  The center of the tank is at (x,y) = (19.2 cm, 18.9 cm).}\label{reconvel}
\end{figure}

The assumption above is clearly not valid close to the source tees,
where the flow is 3D. However, close to the top of the tank the
assumption becomes valid as the flow is quasi-2D. To test the
reconstruction and the validity of the 2D approximation, we compare
the original measured velocity fields to velocity fields
reconstructed from the vorticity fields. A region of an original
velocity field and the reconstructed field is shown in
Fig.~\ref{reconvel}. The RMS difference in the magnitudes of the
original and reconstructed fields is about 2\%. The reconstruction
calculation does well in regions where there is a strong uniform
flow. It does less well where there are large gradients in the
velocity, in particular near vortices. However, on the whole the
reconstructed velocity field follows the same behavior as the
original field.

\section{Velocity and vorticity fields}
\label{flowfields}

\subsection{Transition to quasi-2D flow}

Near the forcing sources (tees) at the bottom of the tank, the flow
is very turbulent (Reynolds number $\approx 6 \times 10^4$) and
three dimensional (Rossby number $\gg 1$).  However, moving
vertically away and up the tank from the forcing, the turbulent
velocities decay. The Reynolds number near the top of the tank where
our data were collected is order 1000, based upon the RMS velocities
of 2 cm/s and typical structure size of 5 cm (estimated from the
velocity correlation function). The relative influence of rotation
also becomes larger so that the Rossby number becomes about 0.3.

\begin{figure}
  \includegraphics[width=83mm]{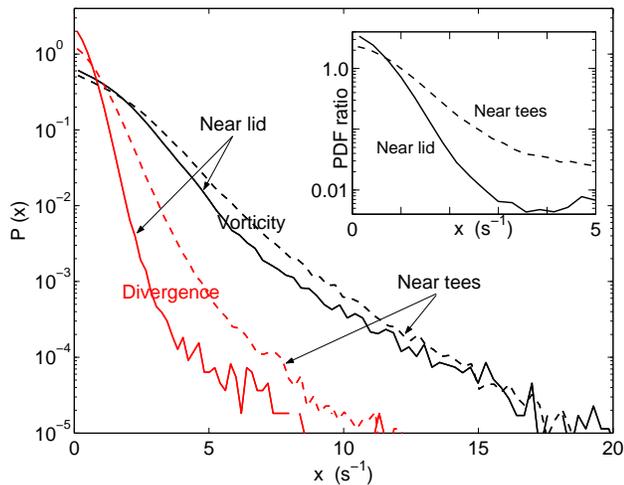}\\
  \caption{(color on-line) The PDF of the magnitude of vertical vorticity and divergence
  at two different heights in our tank, 4 cm below the lid (solid curves) and
  10.4 cm above the top of the tees (dashed curves). Inset: the ratio of the
  divergence PDF to the vorticity PDF, showing the increase
  in the relative magnitude of the divergence near the forcing.}\label{2dness}
\end{figure}

\begin{figure*}
\begin{center}
 \includegraphics[width=\linewidth]{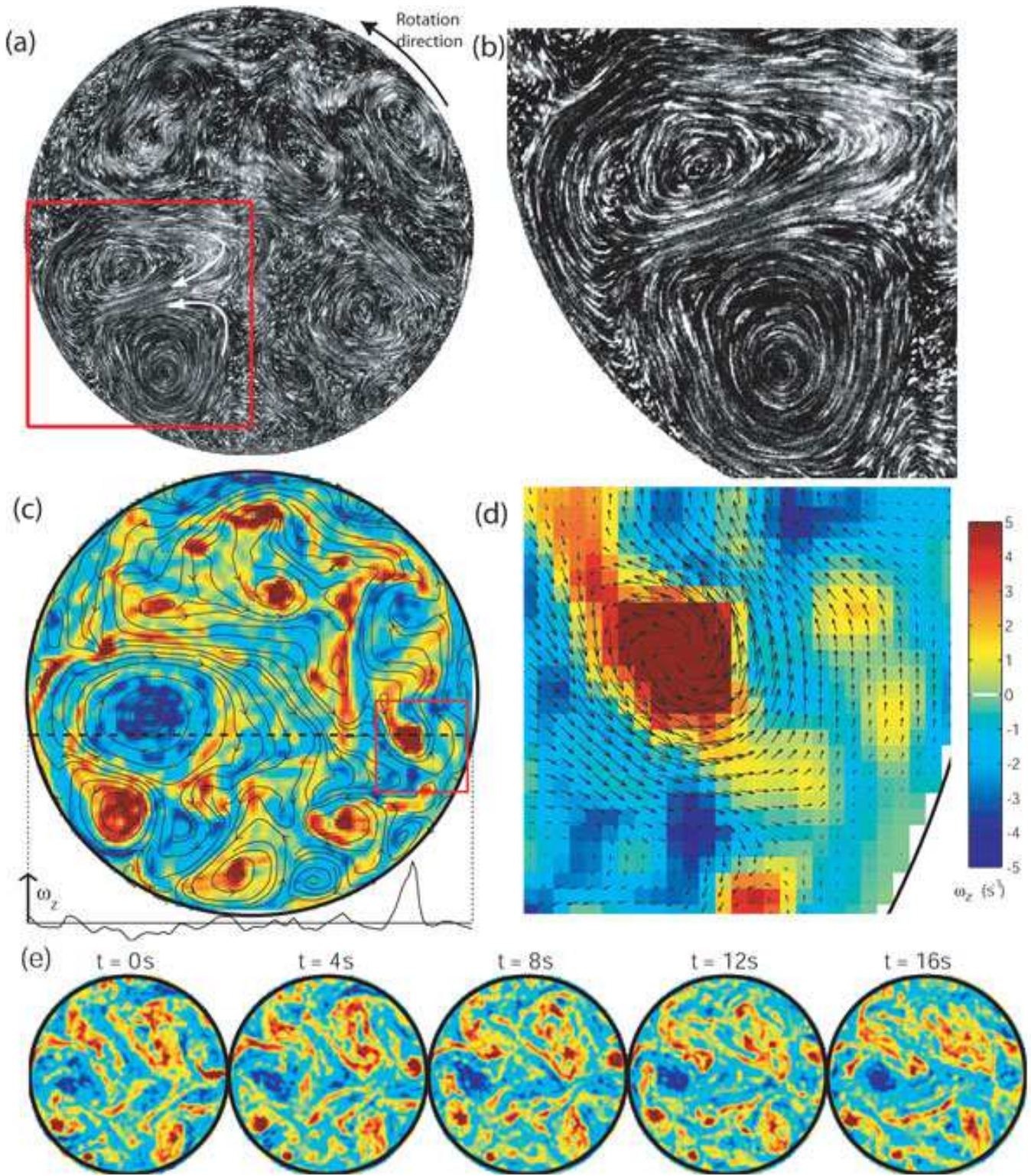}
\end{center}
  \caption{(color on-line) (a) [MOVIE~\cite{particlemovie}] Streak photo (400 ms exposure time) of the particle image fields in a
  horizontal plane 4 cm below the lid of the tank.
  (b) Close-up of the boxed region
  (18.7 cm $\times$ 18.7 cm) showing a cyclone (circular closed particle streaks) and an
  anti-cyclone (elliptical closed particle streaks to the upper-left
  of the cyclone).
  (c) Vorticity field $\omega_z$ with the value along the dashed line shown in the trace
  below the field. The lines indicating the flow structure would be streamlines if the velocity field
  were divergence free, which it is not. Arrows indicate flow direction.
  (d) Close-up of velocity and vorticity
  fields in a 8 cm $\times$ 8 cm region near the boundary. The longest velocity vector corresponds to 6.7 cm/s. The
  vorticity is indicated by the color map, where vorticity values are clipped at 5 s$^{-1}$ to
  render visible weaker structures; the vorticity in the core of
  the strongest cyclone is 22 s$^{-1}$.
  (e) [MOVIE~\cite{vortmovie}] A sequence showing the time evolution of the vorticity field.}\label{vortvelstreak}
\end{figure*}

Fig.~\ref{2dness} compares the magnitude of the divergence and
vertical vorticity of the flow in our tank at two different heights.
The divergence in our flow fields is small relative to the vorticity
and $2 \Omega$ ($\approx$ 5 rad/s). The divergence field consists of
small length-scales near the top and has a weak correlation with the
vorticity field. The ratio of the RMS divergence to the RMS
vorticity is 0.2, of the same order as the Rossby number. Near the
bottom forcing, however, the divergence field becomes larger in
amplitude, length-scale, and more strongly correlates with structure
in the vorticity field. The inset in Fig.~\ref{2dness} shows the
ratio of the divergence to vorticity increases near the forcing,
where the flow is more 3D and decreases near the top, where the flow
is more 2D.

The particles in a plane near the top of the tank are confined
primarily to the plane, and the particle streaks follow persistent
coherent jets, cyclones (a vortex with rotation in the same sense
as the tank), and anti-cyclones without crossing, similar to a 2D
streamline flow [Fig.~\ref{vortvelstreak} (a) and (b)]. In
contrast, the motions of the particles near the bottom of the tank
are much less organized; particles pass through the plane
frequently and the particle streaks can cross and exhibit small
fluctuations due to increased three-dimensionality and turbulence.

\subsection{Persistent coherent structures}

Localized coherent structures including large compact regions of
intense vorticity and wispy filaments of intense vorticity are
evident in Fig.~\ref{vortvelstreak} (c) and (d).  We observe more
cyclones (rotation in the same sense as the tank) than
anti-cyclones, in accord with observations in previous experiments
on rotating turbulent flows~\cite{verdiere.80,hopfinger.82}. The
amplitudes of vorticity at the core of the intense cyclones are
typically more than ten times that of the surrounding flow [see
Fig.~\ref{vortvelstreak} (d) for a close-up of an intense
cyclone]. The anti-cyclones, however, have a typical core
vorticity amplitude only a few times that of the background.

The cyclones, anti-cyclones, and vorticity filaments are
long-lived and are active dynamically. The vortices and vortex
filaments can be tracked by eye as the flow evolves in time
[Fig.~\ref{vortvelstreak} (e)]. Vortices are shed from the wall
and travel across the tank and interact with one another. Vortex
filaments occasionally peel off of vortices, advect with the flow,
and sometimes roll up to form new vortices. Structures may
disappear by merging with other structures or by stretching in a
jet region between opposite sign vortices. Many of the small
intense vortices with very short turn around times ($ < 1$ s) live
on the order of a characteristic decay time ($\sim 10$ s) before
shearing apart or merging with neighboring structures.

Additionally, we see persistent structures such as the large
cyclone (red) and anti-cyclone (blue) that appear in
Fig.~\ref{vortvelstreak}. Such structures often appear in a
preferred location in the tank. Presumably such preferred
locations exist because of inhomogeneities in the forcing by the
tees [cf. Fig.~\ref{apparatus} (c)]. A vortex can be kicked off
its preferred location by a large perturbation when it interacts
with neighboring vortices. If it begins to wander around the tank,
it will generally disappear within a few characteristic decay
times unless it returns to the preferred location. If a structure
moves off its preferred location and disappears, a new structure
will typically form, replacing the pre-existing one within a few
decay times. Some long-lived coherent structures occasionally
persist throughout the entire duration of an experimental run
($\sim 5000$ vortex turnover times). Further, we observe
persistent structures as low as 5 cm above the tees. Observations
closer to the tees are difficult because the three-dimensional
turbulent flow rapidly moves particles into and out of the laser
sheet.

A wide range of spatial scales is visible in the velocity and
vorticity fields [Fig.~\ref{vortvelstreak} (c) and (d)]. The
largest features, approximately 10 cm in size, are coherent
vortices that have a much larger amplitude than their surrounding
region. There are also large amplitude vortex filaments that
stretch up to ~10 cm; these can be as thin as the grid resolution
in the transverse direction.  Small scale persistent structures
less than 1 cm in size yet large in amplitude are also observed.
The region between the various large amplitude coherent structures
is occupied by relatively low amplitude vorticity [see the
vorticity trace in Fig.~\ref{vortvelstreak} (c)].

\section{Decomposition of the vorticity field}
\label{decomposition}

\subsection{Wavelets and wavelet packets}

The discrete wavelet transform (DWT) is a multi-resolution analysis
that successively decomposes the signal into coefficients which
encode coarse and fine details at successively lower
resolution~\cite{wickerhauser}. The basis elements of the transform
$\psi_{s,p}$ correspond to dilations and translations of a mother
wavelet function $\psi$, where $s$ is the scale (dilations) of the
wavelet and $p$ its position (translations). Successive levels of
the transform continue to split the coarse detail coefficients,
effectively analyzing the signal at coarser and coarser resolution.

The discrete wavelet \textit{packet} transform (DWPT) is a
generalization of the discrete wavelet transform. The basis elements
of the wavelet packet transform include, in addition to dilations
and translations, spatial modulation of the mother wavelet at
different resolutions. The basis elements $\psi_{s,p,k}$ take on an
additional parameter $k$, which roughly corresponds to the
modulation of the wavelet packet. In contrast to the DWT, the choice
of basis of the DWPT is not unique \cite{daubechies}. The wavelet
basis is contained within the possible choices of wavelet packet
bases. To select the particular basis to use, a natural choice is
the wavelet packet basis into which the coefficients of the
transform most efficiently represent the signal. This is known as
the ``best basis''. The best basis is typically calculated based
upon the minimization of an effective entropy measure of the
coefficients \cite{wickerhauser}, thus minimizing the ``information
cost'' of the coefficients in the best basis. The flexibility in
basis choice of the DWPT allows the transform to adapt the basis to
the particular signal being analyzed. For example, if the signal
contains regions of rapid fluctuations, the basis choice will
reflect that by including more basis elements with high modulation.

We use the DWT [$O(N)$ operations] and DWPT [$O(N \log_2 N)$
operations] on our experimentally obtained vorticity fields, as
previous authors have done using 2D turbulent flow data from
numerical simulations in a periodic square domain
(e.g.~\cite{farge.92,weiss.97,farge.99}). We use the Matlab wavelet
toolbox and the coiflet 12 (coif2 in Matlab notation) as the
analyzing wavelet. The coiflet family of wavelets (see
reference~\cite{daubechies}, p. 258) has both compact support and
can generate an orthogonal basis. These two properties allow one to
select the localized features of the coherent structures and to
treat the decomposition of the vorticity field as two orthogonal
components. The choice of wavelet does not alter the results
significantly as long as the basis is sufficiently
smooth~\cite{farge.99}.

\subsection{Coherent structure extraction}

We use an algorithm for coherent structure extraction based upon a
de-noising algorithm (e.g. chapter 11 of
reference~\cite{wickerhauser}). In the de-noising, the assumption is
that the original signal can be represented by a few large amplitude
coefficients of an orthogonal transform using an appropriate set of
basis functions, while the noise is contained in the many remaining
coefficients of small amplitude. The de-noising is then performed by
applying a threshold to the resulting coefficients of the transform.
Coefficients above the threshold amplitude are assumed to correspond
to the signal while coefficients below the threshold correspond to
the ``noise''. The few large amplitude coefficients are called the
``coherent'' coefficients, while the many small amplitude
coefficients are ``incoherent''. The coherent and incoherent parts
of the vorticity are then reconstructed by the inverse transform.

The choice of the threshold separating the coherent and incoherent
parts of the signal is based upon a measure of the number of
significant coefficients $N_o$, which is the theoretical dimension
of the signal, defined by

\begin{equation}\label{sigcoef}{N_o(f)} = {e^{H(f)}}\end{equation}

\noindent where $H(f)$ is the entropy of a discrete signal $f =
[f_i]$ (where $[f_i]$ is the set of discrete values of an arbitrary
signal $f$). In our case, the $f_i$'s correspond to the discrete
values of our vertical vorticity measurements. The entropy $H(f)$ is
defined as

\begin{equation}\label{entropy}
{H(f)} = {-\sum_{i=1}^N p_i \log{p_i}}\end{equation}

\noindent where $p_i  = |f_i|^2 / ||f||^2$ is the normalized square
modulus of the $i^\mathrm{th}$ element of the signal, with $N$ the
number of elements and ${||f||^2} = \sum_i |f_i|^2$.  $N_o$
indicates how many of the largest coefficients should be retained to
give an efficient, low entropy, representation of the signal.

Thus, the decomposition algorithm is the following. We take the
transform of an individual measured vorticity field. For the DWPT,
the best basis is used. Then we find the number of significant
coefficients $N_o$ of the transformed vorticity field in the
transform basis. The threshold is therefore based upon the value of
the $N_o^{th}$ largest coefficient. Coefficients whose modulus is
larger (smaller) than the threshold correspond to the coherent
(incoherent) part of the vorticity field. We then take the inverse
transform to get the coherent part in physical space. The incoherent
field is reconstructed by subtracting the coherent from the total
field. The process is then repeated for each of our vorticity
fields. Our algorithm contains no adjustable parameters other than
the selected wavelet and is based upon the assumption that the flow
field has a low entropy component, corresponding to the coherent
structures, and a high entropy component, corresponding to an
incoherent background.

Our algorithm is different than the various iterative algorithms of
Farge et al.~\cite{farge.99} and
others~\cite{wickerhauser,weiss.97}. In the de-noising algorithm
described in \cite{wickerhauser,weiss.97}, successive iterations
extract a fraction of the coherent coefficients until a stopping
criterion is met. The fraction of retained coefficients at each step
and stopping criterion are adjustable parameters. However, the basis
of the DWPT can change at each iteration. The number of coefficients
retained therefore loses its meaning since coefficients can be
selected from completely different bases. Therefore, we only perform
a single iteration without changing the basis for the DWPT. In the
algorithm of Farge~\cite{farge.99}, an a priori assumption is made
of Gaussian white noise incoherent component superimposed upon a
non-Gaussian component of coherent vortices. The threshold is then
iteratively found to separate the two without adjustable parameters.

\subsection{FFT, JPEG compression, and Okubo-Weiss based techniques}

To examine the relative efficiency of the wavelet-based
decompositions, we compare the DWPT and DWT to three other
algorithms; one based upon a Fourier transform, a second on basic
JPEG compression, and a third on the method developed by Okubo
\cite{okubo.70} and Weiss \cite{weiss.91}.

For the Fourier transform algorithm we threshold the coefficients
of a 2-dimensional fast Fourier transform (FFT) of our vorticity.
The coherent fields are then constructed via an inverse FFT of the
largest amplitude coefficients. The incoherent remainder fields
are constructed by subtracting the coherent fields from the
original fields.

The basic JPEG algorithm consists of subdividing an image into
$8\times8$ blocks and taking the discrete cosine transform over the
sub-blocks. The coefficients are arranged based upon the global
average of the amplitudes among the sub-blocks. The average of the
amplitudes is used to determine which modes to keep. After the
threshold, each sub-block retains the same number of modes, each in
the same position. Thus the number of modes retained must be a
multiple of the number of sub-blocks which compose the image (for
our $128\times128$ fields, this is 256). For more information on the
JPEG image compression standard see reference~\cite{jpeg2000book},
chapter 3.

The Okubo-Weiss criterion splits the fields into elliptic and
hyperbolic regions, dominated by vorticity and strain respectively.
The regions dominated by vorticity and strain are then separated and
taken to be the coherent and incoherent fields. For details on the
Okubo-Weiss criterion see reference~\cite{basdevant.94}.


\section{Results}
\label{results}

\begin{figure*}
  \includegraphics[width=\linewidth]{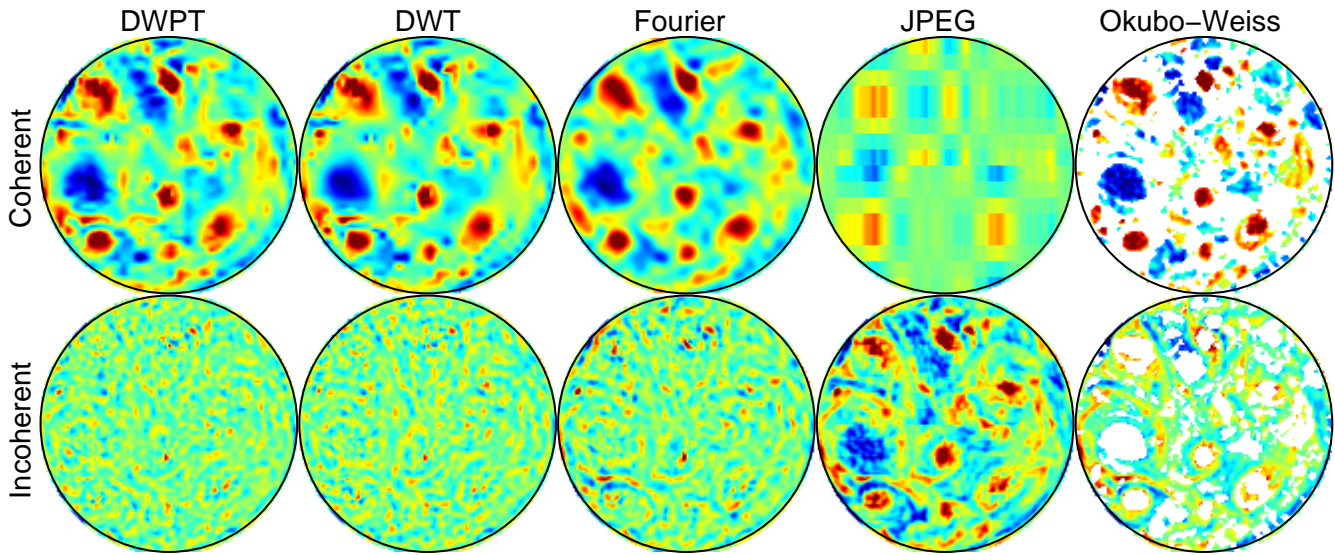}\\
  \caption{(color on-line) Decomposed vorticity fields, TOP: Coherent BOTTOM: Incoherent remainder fields,
  all for the same number of coefficients retained (except for Okubo-Weiss and approximate for JPEG).
  The colormap corresponds to the intensity of vorticity and is the same for all images and the same as used in
  Fig.~\ref{vortvelstreak}. [MOVIE~\cite{decmpfieldsmovie}]} \label{vortfields}
\end{figure*}

\subsection{Decomposed vorticity fields}
\label{decmpvortfields}

The resulting coherent and incoherent vorticity fields obtained from
the wavelet packet, wavelet, Fourier, and JPEG techniques are
compared in Fig.~\ref{vortfields}. The Okubo-Weiss criterion was
also tested for comparison with the other decomposition methods. The
fields shown are constructed with the same number of coefficients
for the DWPT, DWT, and Fourier transforms, and the closest
approximate number for JPEG.

The coherent fields in Fig.~\ref{vortfields} appear to retain the
large scale coherent structures [cf. Fig.~\ref{vortvelstreak}(c)
and (e)]. The DWPT, DWT, and Fourier preserve the structure of the
total vorticity field. The Okubo-Weiss criterion excises primarily
the regions of large amplitude vorticity, which generally
corresponds to the cores of vortices. However, it extracts only
the cores and leaves behind the peripheries. The JPEG does poorly
with so few coefficients; it barely picks out some of the stronger
vortices in the field.

For the same number of coefficients, 2.4\%, the DWPT and DWT do a
better job of extracting structure in the fields than the Fourier
decomposition, which leaves more structure behind in the incoherent
field (Fig.~\ref{vortfields} [MOVIE~\cite{vortmovie}]). The JPEG
incoherent field is large amplitude because of the few coefficients
retained in the coherent field. The incoherent fields resulting from
the DWPT, DWT, and the Fourier are mostly devoid of large-scale
structures and are much smaller in amplitude. The incoherent fields
are also poorly correlated in time and are devoid of large scale
structures or any other feature which can be tracked by eye
(Fig.~\ref{vortfields} and associated movie).

The Okubo-Weiss criterion selects the centers of the vortices where
vorticity dominates strain. This accounts for a large portion of the
enstrophy. It extracts about 74\% of the enstrophy in the vorticity
dominated regions, which cover about 40\% (not coefficients) of the
flow. However, the method leaves behind holes in the vorticity field
that act as coherent structures and are often surrounded by large
values of vorticity. The resulting ``incoherent'' field thus retains
coherent properties. The Okubo-Weiss criterion also does poorly in
recovering the statistics of the total field, as shown in Table
\ref{table1}. Furthermore, the decomposition shown in
Fig.~\ref{vortfields} was performed without regard to the
Okubo-Weiss validity criterion, which restricts the application of
the criterion to the very centers of the vortices or regions of very
large vorticity \cite{basdevant.94}. If the validity criterion is
applied, much less is retained in the coherent field, and the
incoherent field ends up with large values of vorticity.

\begin{table*}
\begin{footnotesize}
\begin{tabular}{ c c c c c c c c c c c c c c}
   & Decomposition &  & \multicolumn{2}{c}{Wavelet packet}  &  \multicolumn{2}{c}{Wavelet} &  \multicolumn{2}{c}{Fourier} &  \multicolumn{2}{c}{JPEG} & \multicolumn{2}{c}{Okubo-Weiss}&\\ 
   & Quantity                               & Total & \multicolumn{2}{c}{coherent-incoherent}  & \multicolumn{2}{c}{coherent-incoherent} & \multicolumn{2}{c}{coherent-incoherent} & \multicolumn{2}{c}{coherent-incoherent} & \multicolumn{2}{c}{coherent-incoherent}\\ \hline
   & coefficients retained  (\%)            &  100 & 2.4  & 97.6 & 2.7  & 97.3 & 4.1  & 95.9 & 13.1 & 86.9 & 40.2  & 59.8  & \\
   & Enstrophy/field  (s$^{-2}\times 10^4$) & 4.33 & 3.63 & 0.63 & 3.67 & 0.62 & 3.68 & 0.56 & 2.92 & 1.33 & 3.21  & 1.12  & \\
   & Vorticity skewness                     & 0.78 & 0.80 & 0.02 & 0.85 & 0.04 & 0.62 & 0.20 & 0.53 & 0.24 & 1.22  & -0.07 & \\
   & Vorticity kurtosis                     & 7.09 & 7.03 & 3.88 & 7.58 & 3.42 & 5.37 & 7.85 & 5.27 & 6.55 & 11.81 & 8.28  & \\
\end{tabular}
\end{footnotesize}
  \caption{Statistical properties of the decomposed vorticity fields using entropy criterion.}\label{table1}
\end{table*}

\begin{figure}[h]
  \includegraphics[width=\linewidth]{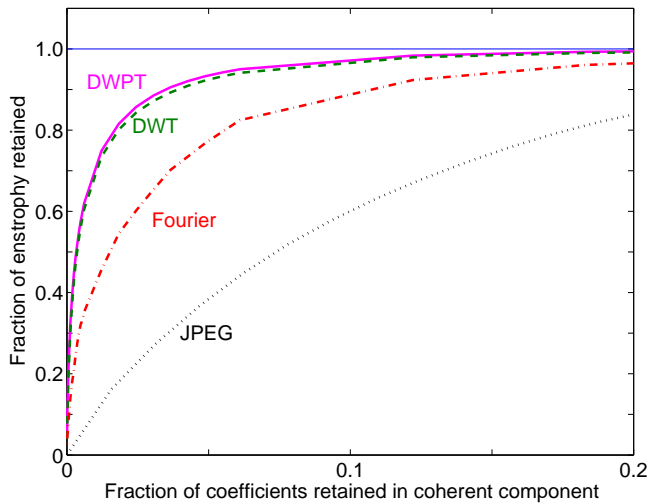}\\
  \caption{(color on-line) The percent enstrophy retained
  in the largest amplitude coefficients as a function of the number of coefficients kept.
  In order from best (most efficient) to worst: wavelet packet, wavelet, Fourier, JPEG.}\label{compressioncurve}
\end{figure}

\begin{figure}[h]
  \includegraphics[width=\linewidth]{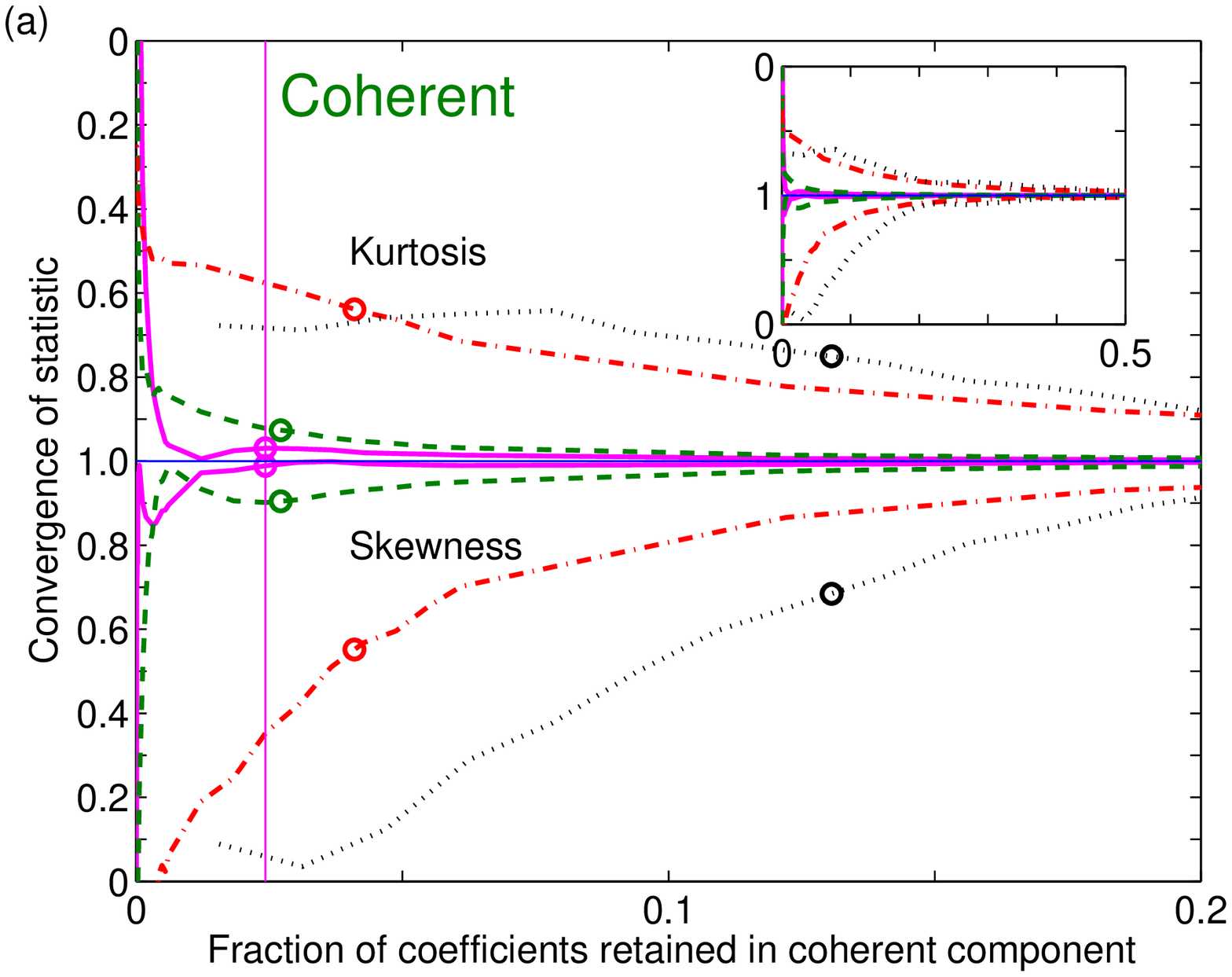}\\
  \includegraphics[width=\linewidth]{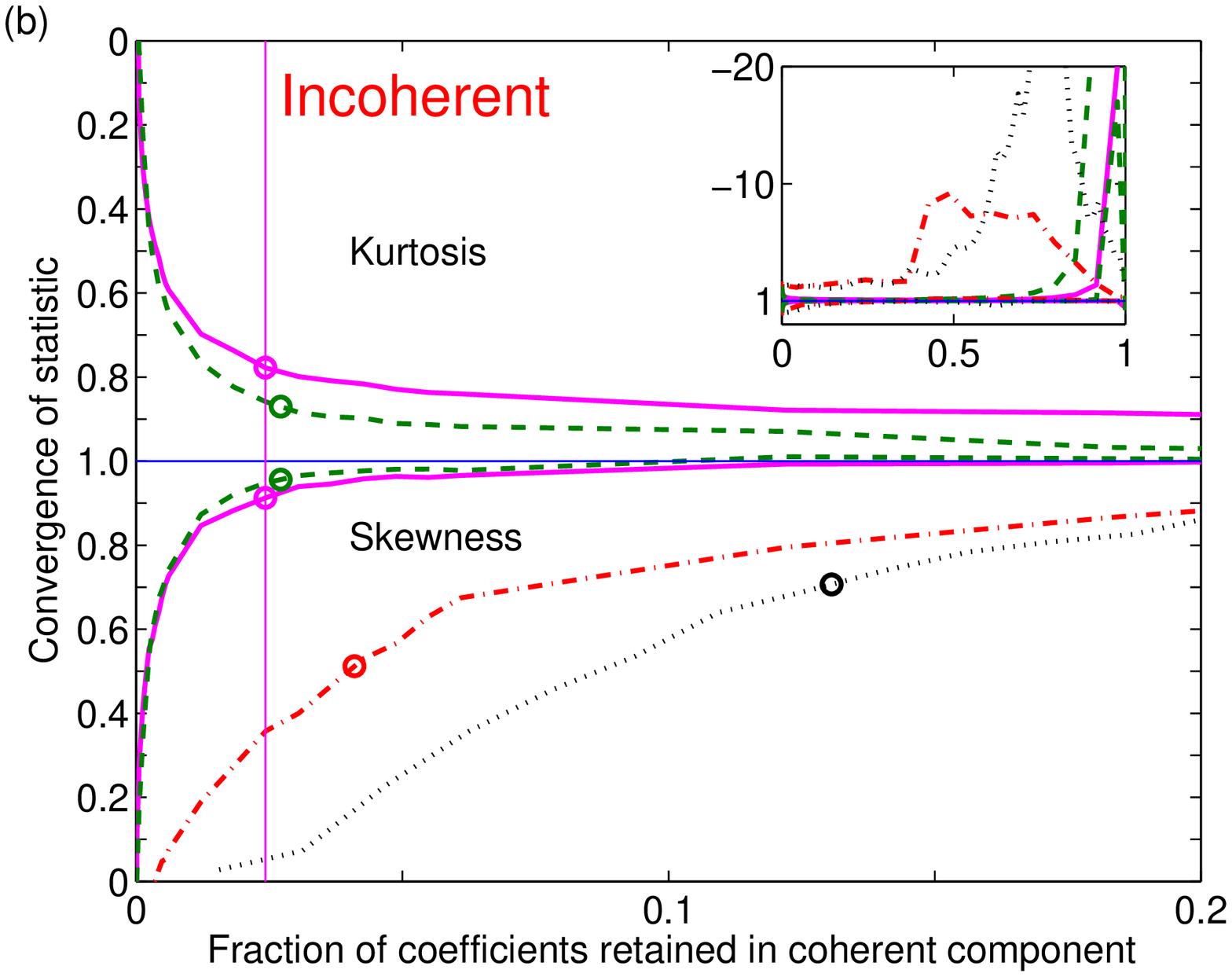}\\
  \caption{(color on-line) Convergence of the skewness and kurtosis of the vorticity PDFs
  for the various decompositions, plotted with quantities defined so that they approach unity
  as they converge. The vertical line indicates the fraction of
  coefficients retained in the wavelet packet decomposition.
  (a) Convergence of the coherent component. Plotted: kurtosis ratio: $|k_{\mathrm{coh}}/k_{\mathrm{total}}|$,
  skewness ratio: $|s_{\mathrm{coh}}/s_{\mathrm{total}}|$.
  (b) Convergence of the incoherent component. Plotted: kurtosis ratio: $2-k_{\mathrm{incoh}}/3$,
  skewness ratio: $1 - |s_{\mathrm{incoh}}/s_{\mathrm{total}}|$. Note that unity on the graph represents
  the respective statistic for a Gaussian distribution: skewness = 0, kurtosis = 3.
  The kurtosis of the Fourier and JPEG incoherent field is large and off the main plot.
  The insets show the convergence behavior at large numbers of retained
  coefficients. The open circles on the curves correspond to the fraction of coefficients
  retained for the corresponding techniques using the entropy criteria as a threshold.
  }\label{convergence}
\end{figure}

Statistical properties of the decompositions are shown in
Tables~\ref{table1} and~\ref{table2}. Table~\ref{table1} lists the
results from setting the threshold of the various transforms based
upon the number of significant coefficients calculated from
equation~(\ref{sigcoef}). On average about 2-3\% of the
large-amplitude coefficients of the wavelet-based decompositions
were retained in the coherent fields, which account for about 85\%
of the total enstrophy of the flow. The Fourier retains roughly
the same enstrophy but requires more coefficients, about 4\%. The
JPEG decomposition retains only 67\% of the total enstrophy with
about 13\% of the large-amplitude coefficients.

Owing to the compact support of their basis functions, fewer
coefficients were needed for the DWPT and DWT, based upon the
entropy criterion in equation~(\ref{entropy}), than for the Fourier
and JPEG decompositions. Compression curves for enstrophy of the
various decompositions applied to our vorticity fields are shown in
Fig.~\ref{compressioncurve}. The DWPT and DWT both do similarly well
at extracting the enstrophy in the fields with only a small number
of coefficients. The DWPT probably does slightly better due to the
adaptability of its basis. The Fourier based methods do not converge
as rapidly as the wavelet-based methods. This is likely due to the
non-locality of the basis functions when applied to a field which
has localized structures and sharp features.

The wavelet-based algorithms also do a better job preserving the
skewness and kurtosis of the total vorticity in the coherent
component, while the incoherent components are much closer to a
Gaussian distribution (skewness = 0, kurtosis = 3).
Table~\ref{table2} displays the results from setting the threshold
on the transforms so that each method retains 2.4\% of the
coefficients in the coherent component (since the coefficients
JPEG must be a multiple of the number of $8\times8$ blocks, 256,
this restriction is relaxed). For the same number of retained
coefficients, the DWT and DWPT clearly outperform the Fourier
method in terms of preserving the statistics of the total
vorticity field.

Figure~\ref{convergence} shows that the coherent fields from the
wavelet packet and wavelet decompositions rapidly converge to the
statistics of the total vorticity field. The coherent fields
preserve the non-Gaussianity of the vorticity PDF without having to
extract many coefficients. The respective incoherent fields also
rapidly converge to Gaussian statistics. This suggests that wavelets
and wavelet packets do a good job of extracting the coherent
structures from the vorticity field. The near-Gaussianity of the
incoherent field suggest that the transforms have left the remainder
without coherent structure.

The coherent components of the Fourier and JPEG decompositions,
however, converge to the statistics of the total fields much more
slowly. To obtain the same level of fidelity, both the Fourier and
JPEG must extract many more coefficients. This inefficiency would
result in over-transformed fields. Further, the insets in
Fig.~\ref{convergence} (b) show that the kurtosis of the
incoherent component never converges to anything small, but
increasingly deviates from Gaussian. This suggests that the
Fourier decomposition is not separating the coherent structures
from the background.

\begin{figure}
  \includegraphics[width=83mm]{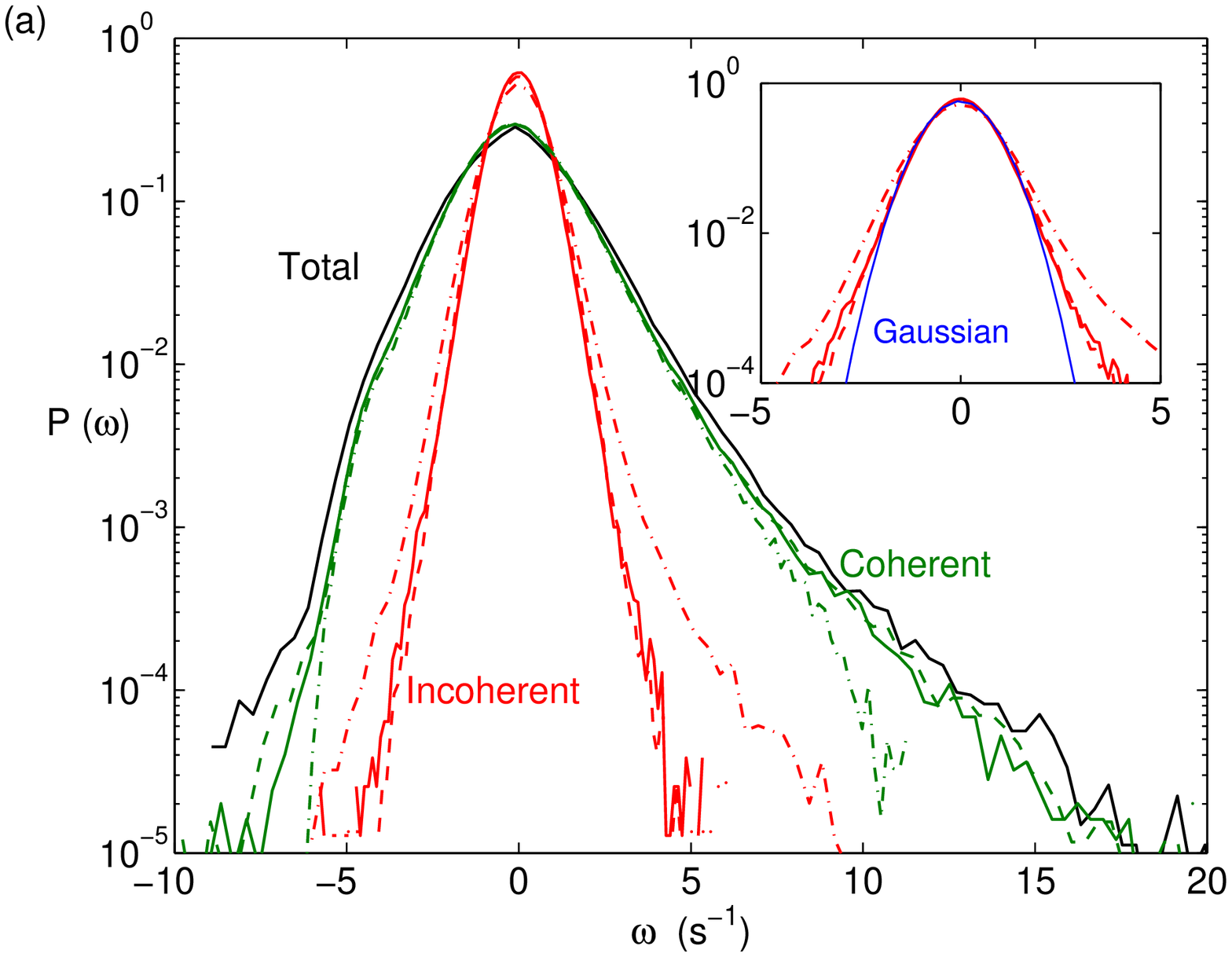}
  \includegraphics[width=83mm]{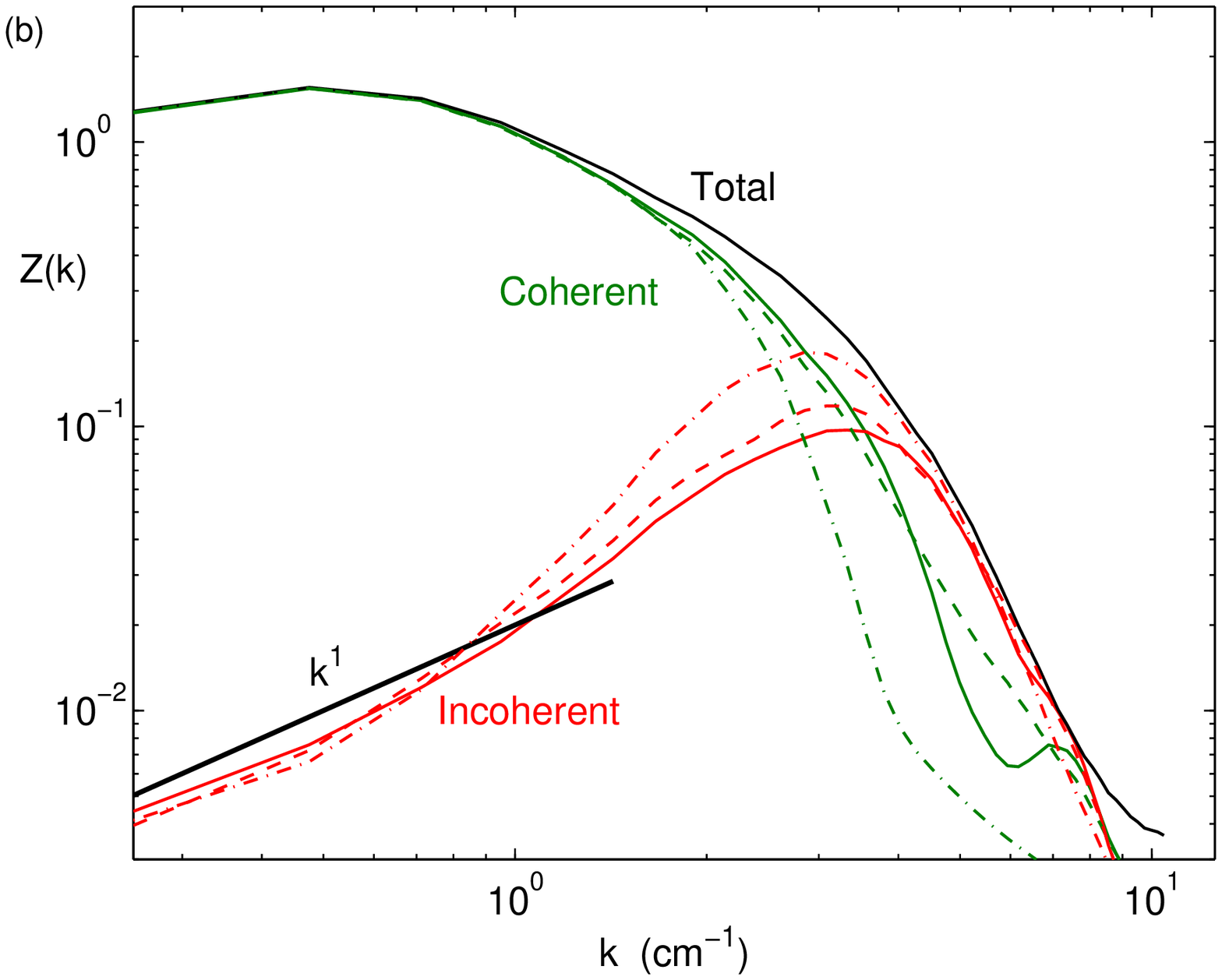}\\
  \caption{(color on-line) (a) Vorticity PDFs of the total field, coherent field, and
    incoherent field, where the inset shows the incoherent PDFs
    plotted together with a Gaussian fit to the wavelet packet
    results. (b) Enstrophy spectra Z(k) of total, coherent, and incoherent
    fields, calculated on a square subsection of our circular
    domain. The solid, dashed, and dash-dotted lines correspond to the
    wavelet packet, wavelet, and Fourier decompositions respectively.
    The k$^1$ line is the spectral slope for Gaussian white noise in 2D.
    The wavenumber is defined as k  = $2\pi/L$.}\label{pdfspec}
\end{figure}

\begin{table*}
\begin{footnotesize}
\begin{tabular}{ c c c c c c c c c c c c }
   & Decomposition &  & \multicolumn{2}{c}{Wavelet packet}  &  \multicolumn{2}{c}{Wavelet} &  \multicolumn{2}{c}{Fourier} &  \multicolumn{2}{c}{JPEG}&\\ 
   & Quantity                               & Total & \multicolumn{2}{c}{coherent-incoherent}  & \multicolumn{2}{c}{coherent-incoherent} & \multicolumn{2}{c}{coherent-incoherent} & \multicolumn{2}{c}{coherent-incoherent}\\ \hline
   & coefficients retained  (\%)            &  100 & 2.4  & 97.6 & 2.4  & 97.6 & 2.4  & 97.6 &  3.0 & 97.0 &   \\
   & Enstrophy/field  (s$^{-2}\times 10^4$) & 4.33 & 3.63 & 0.63 & 3.60 & 0.69 & 3.28 & 0.92 & 0.96 & 3.18 &   \\
   & Vorticity skewness                     & 0.78 & 0.80 & 0.02 & 0.86 & 0.04 & 0.51 & 0.32 & 0.06 & 0.72 &   \\
   & Vorticity kurtosis                     & 7.09 & 7.03 & 3.88 & 7.62 & 3.47 & 4.77 & 7.85 & 4.84 & 7.08 &   \\
\end{tabular}
\end{footnotesize}
  \caption{Statistical properties of the decomposed vorticity fields retaining the same number of modes, 2.4\%, as the DWPT decomposition (approximate for JPEG).}\label{table2}
\end{table*}

The probability distribution function (PDF) of vorticity has broad
wings (Fig.~\ref{pdfspec}), which correspond to the large vorticity
values that occur in the coherent vortices and vorticity filaments.
The preference for cyclonic (positive vorticity) structures over
anti-cyclonic structures appears as a large positive skewness (see
also Table~\ref{table1} or ~\ref{table2}). While the PDF for the
DWPT and DWT coherent vorticity field are nearly the same as that
for the total vorticity field, the PDF for the corresponding
incoherent field is narrower and more symmetric, indicating the lack
of high intensity structures.

Consider the enstrophy spectrum of the vorticity fields, as shown
in Fig.~\ref{pdfspec} (b). It is clear from the enstrophy spectra
of the total field that most of the enstrophy is contained in long
wavelengths (small $k$). This agrees well with the observation of
the vorticity fields and the dominance of large structures with
large amplitude vorticity. The enstrophy spectrum, shown in
Fig.~\ref{pdfspec}, does not contain a well defined scaling region
corresponding to a cascade of enstrophy. A cascade would not be
expected for our quasi-2D flow, which is forced by the broad-band
3D turbulence in the bottom of the tank rather than by injection
of energy at a single well-defined wavenumber. The presence of
large structures which extend well into the 3D region alone tells
us that we must have a broad spectrum energy injection.

\begin{figure*}[h]
  \includegraphics[width=3in]{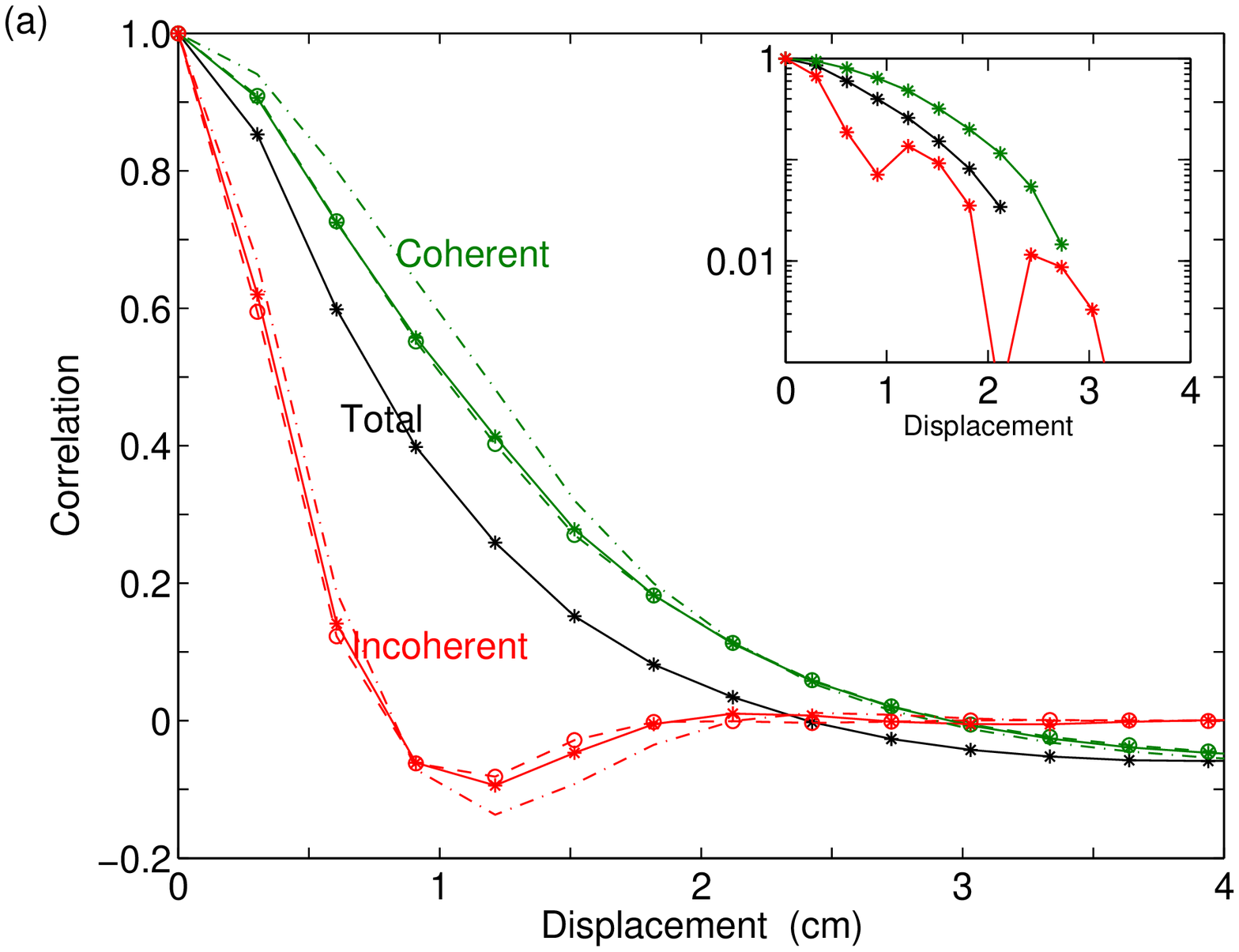}
  \includegraphics[width=3in]{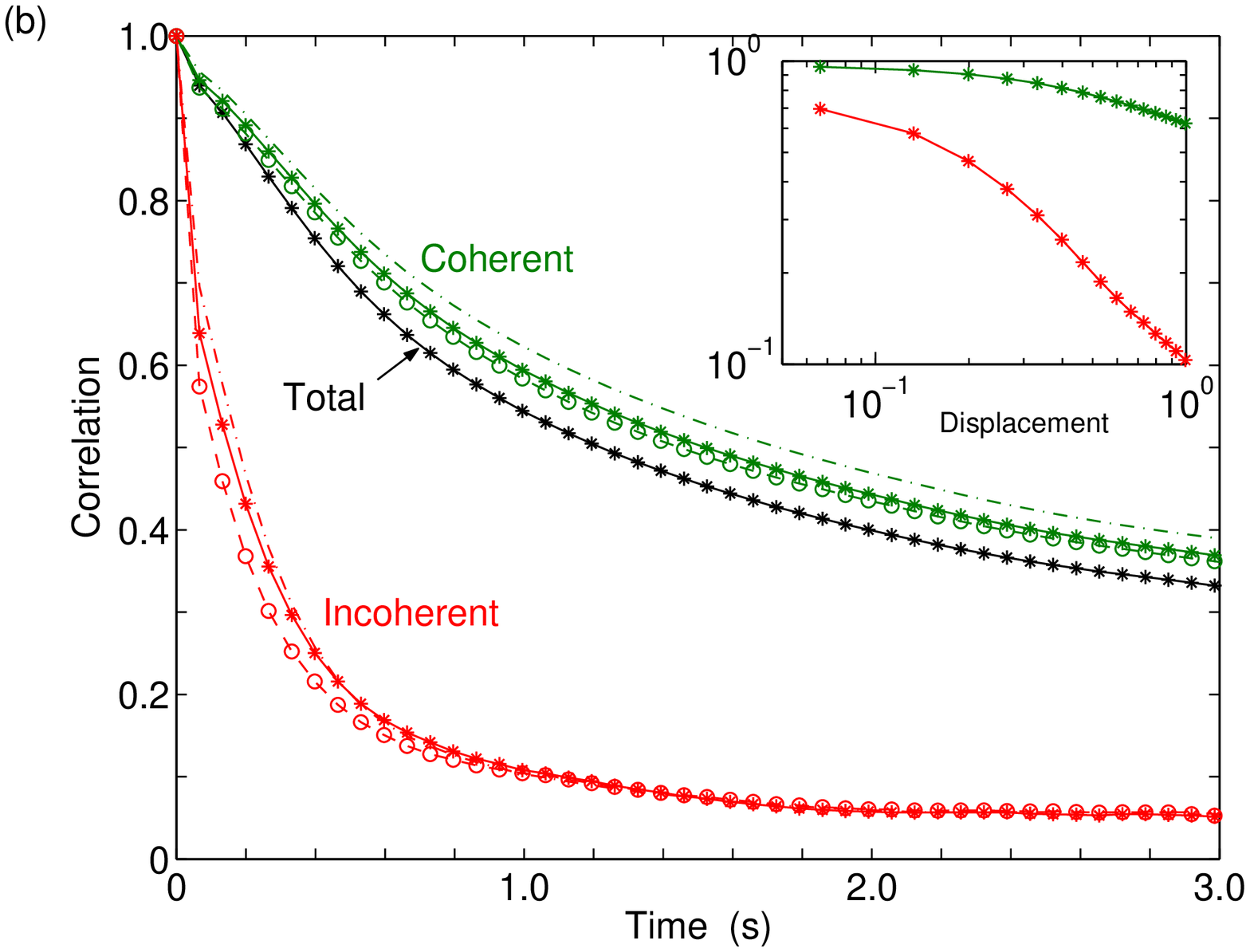}
  \caption{(color on-line) (a) Space and (b) time correlations of the decomposed vorticity fields.
  The solid, dashed, and dash-dotted curves correspond to the DWPT, DWT, and
  Fourier decompositions.}\label{corplots}
\end{figure*}

The coherent field contains the same large structures that are
present in the total field. Accordingly, the enstrophy spectrum of
the coherent field matches that of the total field at long
wavelength [Fig.~\ref{pdfspec} (b)]. The incoherent field contains
negligible enstrophy at large scales, indicating the lack of large
scale features. This is apparent in the vorticity decomposition
(see Fig.~\ref{vortfields}).  The incoherent field has only small
structures and is the dominant contribution to the enstrophy at
large $k$.

The wavelet-based and the Fourier decompositions also retain the
spatial and temporal correlations in the coherent field, as
Fig.~\ref{corplots} illustrates with the DWPT and DWT yielding
almost the same results. The long time correlation is in part due
to the presence of long lived coherent structures. In contrast,
for the incoherent field the spatial and temporal correlations are
short ranged, indicating the absence of large scale and long-lived
structures.

\subsection{Transport of passive scalar particles and concentration fields}

\begin{figure}[h]
  \includegraphics[width=83mm]{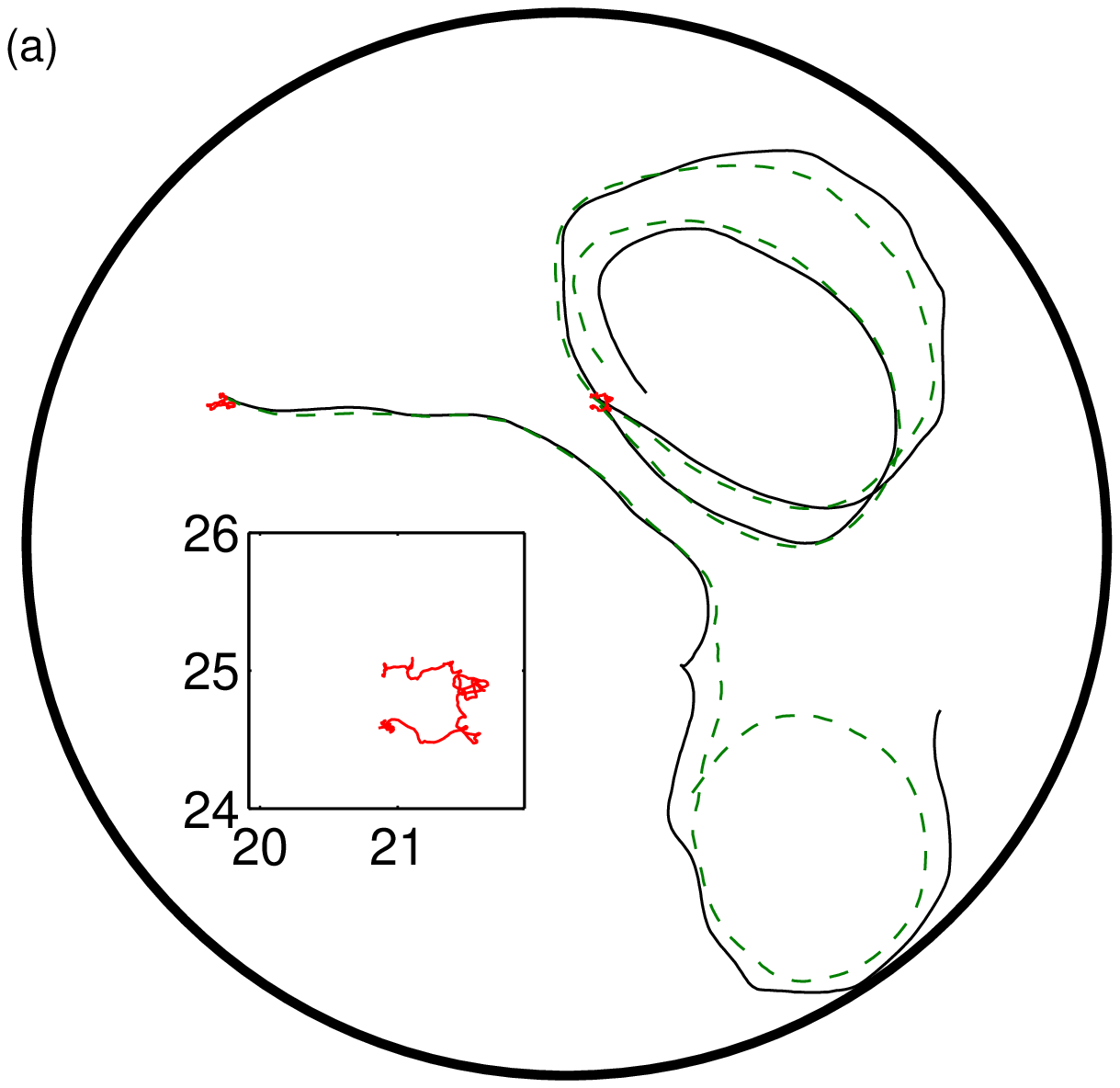}
  \includegraphics[width=83mm]{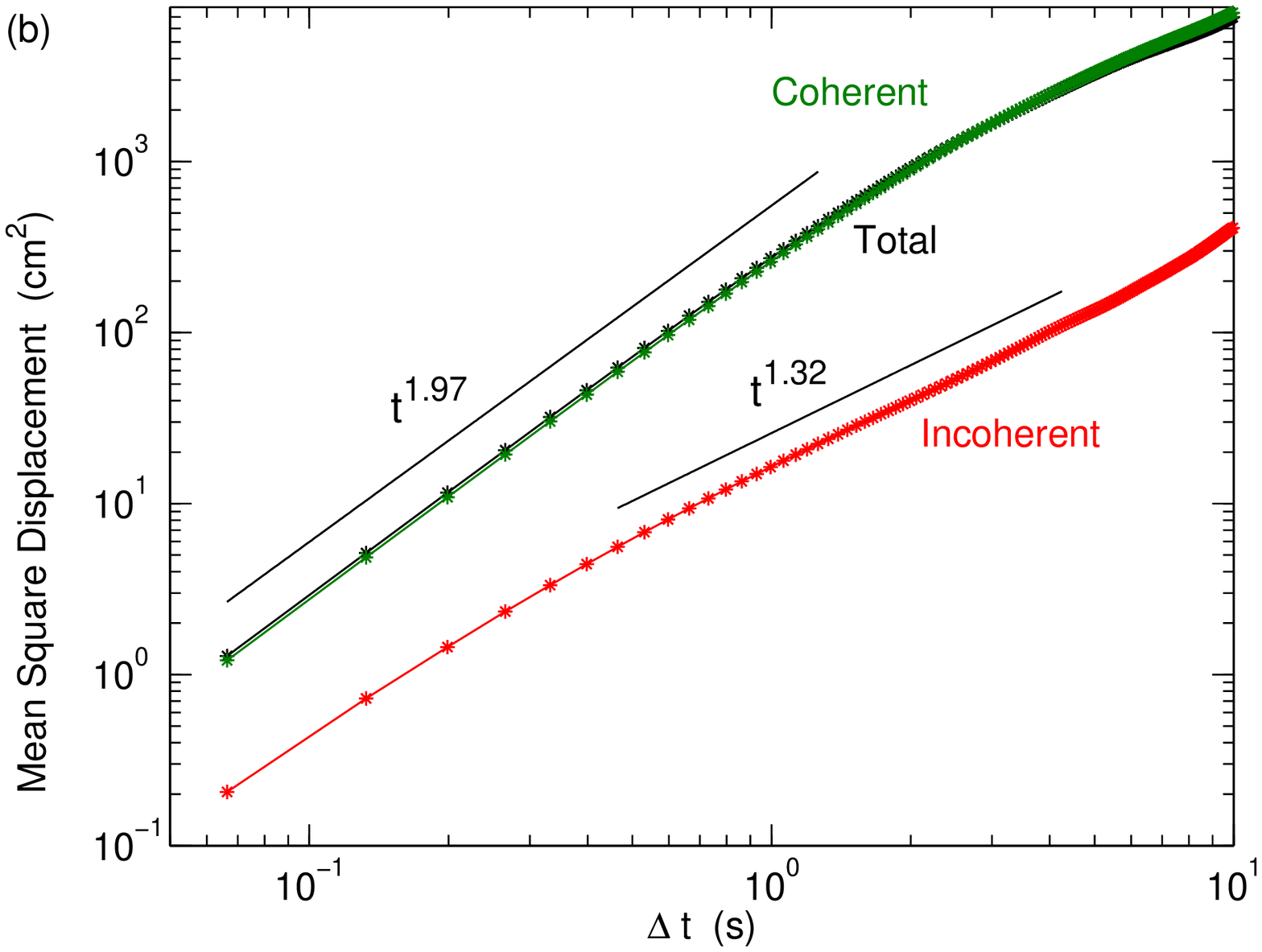}\\
  \caption{(color on-line) (a) Simulated paths of numerical tracer particles in the
  total field (solid curves) and in the coherent (dashed curves) and incoherent field
  (red) for the DWPT (total time 20 s). The inset shows an expanded view
  of the path in the incoherent field.
  (b) Mean squared displacements for the tracers in the total,
  coherent, and incoherent velocity fields. The exponents for
  the total and coherent fields are 1.97 while
  for the incoherent remainder the exponent is 1.32 over the range
  indicated. Both results are for the wavelet packet.}\label{tracerpaths}
\end{figure}

Paths computed for some passive scalar ``particles" in the
measured velocity fields are shown in Fig.~\ref{tracerpaths} (a).
Over the 20 s duration of the measurement, the particles are
advected around a large portion of the tank. The motion of the
particles depends on their location in the flow. For example, a
particle can spend time caught in a vortex, following the motion
of the vortex as it meanders slowly in a localized region of the
tank [path in upper right of Fig.~\ref{tracerpaths} (a)].
Occasionally the particle will escape from a vortex and then be
carried by a high velocity jet, which can transport a particle
large distances in a short time. A particle may then be captured
by a vortex [e.g. path ending in lower right of
Fig.~\ref{tracerpaths} (a)].

The resulting mean squared displacement of the tracer particles in
the velocity fields obtained by experiment is shown in
Fig.~\ref{tracerpaths} (b). It exhibits approximately $t^{2}$
scaling at short times, which looks ``flight-like''
\cite{solomonweeks.93}. At long times the finite size of the tank
becomes significant and the scaling exponent becomes smaller.

Both the wavelet and wavelet packet coherent fields disperse the
tracer particles similarly around the tank; the scaling behavior
of the mean squared displacement, not shown in this paper, is also
the same. Differences in the paths are due to the sensitive
dependence upon initial conditions of a turbulent flow field.

\begin{figure}[h]
  \includegraphics[width=83mm]{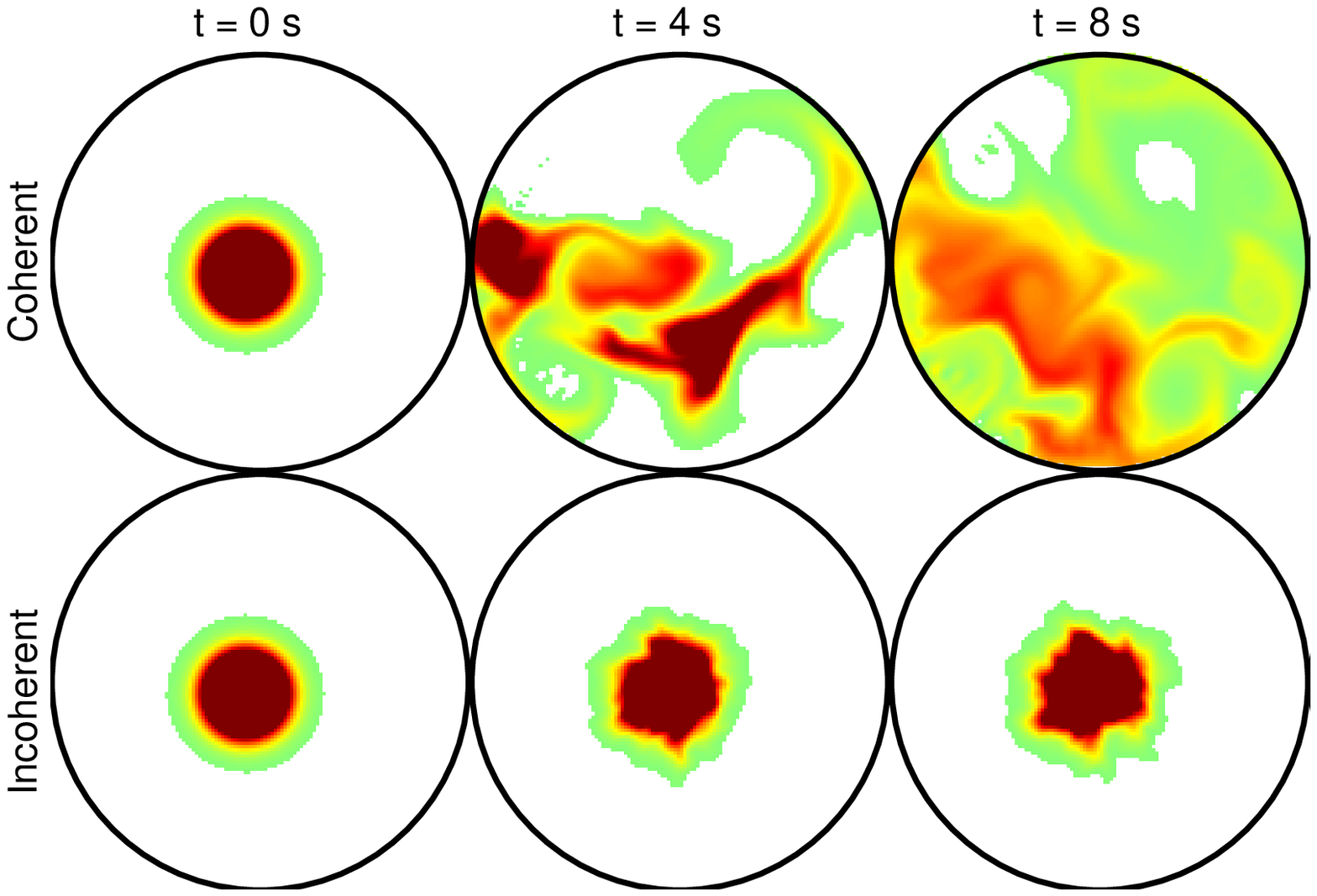}\\
  \caption{(color on-line) [MOVIE~\cite{pscalarmovie}] Advection of a passive scalar field in the velocity field
  of the DWPT, as computed from the advection-diffusion equation~(\ref{advecdiff}).
  By eight seconds the scalar field has been significantly mixed
  by the coherent field, while it appears only to have been diffused in the incoherent field.} \label{pscalaradvection}
\end{figure}

We find a striking difference in the behavior of a numerically
integrated passive scalar particle in the coherent and incoherent
fields.  The incoherent fields make no significant contribution to
the transport properties; rather, particles are confined to a
small region, as shown in the inset in Fig.~\ref{tracerpaths} (a).
This is due to the rapid decorrelation in time of the incoherent
fields. The rapid fluctuations cause the tracers to jiggle around
and the scaling of the mean squared squared displacement of the
tracers in the incoherent fields approaches that of a random walk.

The evolution of the passive scalar field in the measured velocity
field is similar to that for tracer particles. Results for the
coherent and incoherent fields produced by the DWPT are shown in
Fig.~\ref{pscalaradvection}, which illustrates the stretching and
folding; the results for the total field are similar to that of the
coherent field. By four seconds the scalar has already been
significantly stretched by the velocity fields. There is no
significant advection in the incoherent fields and is similar to
results obtained with pure diffusion (no advection term). This is
expected from the short time and space correlations of the
incoherent fields (Fig.~\ref{corplots}). Similar results were
obtained for the DWT and Fourier methods. Our results agree well
with those of Beta et al.~\cite{beta.2003} who, using the DWT, found
that the coherent field is responsible for the mixing in a
numerically simulated 2D turbulent flow.

\section{Discussion}
\label{discussion}

We find that flow which is strongly forced near the bottom of a deep
rapidly rotating tank evolves with increasing height in the tank
from 3D turbulence into rotation-dominated turbulence. This is due
to the competition between the turbulent forcing which is 3D, and
the rotation which tends to two-dimensionalize the flow. Near the
forcing the turbulent velocities are large and the flow is 3D. Away
from the forcing the spatial decay of turbulence allows the rotation
to become increasingly important with increasing height. Motions of
the particles in the flow are more organized in the horizontal plane
near the top, consistent with two-dimensional columnar vortices. The
flow in the rotationally dominated (Rossby number $\approx 0.3$)
region of our tank has a strongly non-Gaussian vorticity PDF and
exhibits properties of 2D turbulence, such as large-scale long-lived
coherent vortices, vortex filamentation, and merger.

The vertical component of vorticity describes the flow well near the
top where the flow is quasi-2D as a consequence of the rotation. We
are thus able to use the projected flow field as measured in a
horizontal plane for analysis with wavelet and Fourier-based
decompositions. We have shown that wavelet based transforms can be
used to separate the coherent structure-containing quasi-2D
turbulent flow into a non-Gaussian coherent component represented by
as few as 3\% of the large-amplitude, low entropy coefficients of
the transform and a nearly Gaussian incoherent component,
represented by the remaining small-amplitude, high entropy
coefficients. We are able to obtain this result without an a priori
assumption of Gaussianity or non-Gaussianity of the two components.

The discrete wavelet packet transform (DWPT) and the discrete
wavelet transform (DWT) yield very similar results, despite the
adaptability of the DWPT basis. The DWT is therefore made preferable
by its faster computation [$O(N)$ versus $O(N \log_2 N)$
operations].

Because our flow fields contain compact structures, the localized
basis functions of the DWPT and DWT outperform Fourier and JPEG
decompositions. Both the DWPT and DWT have more rapid convergence of
the statistics of the extracted coherent component toward that of
the total flow. The rapid convergence of the skewness and kurtosis
of the vorticity PDF suggest that the wavelet based methods have
efficiently captured the coherent structures. The Fourier and JPEG
converge much more slowly and are thus unable to efficiently capture
the large higher moments of the vorticity PDF. This suggests that
the Fourier and JPEG methods have not really extracted the coherent
structures despite the appealingly smooth visual appearance of the
structures in the coherent field of the Fourier. Indeed, the
superior performance of the DWT over the JPEG is exploited by the
emerging next generation JPEG2000 image compression standard, which
uses a bi-orthogonal DWT~\cite{jpeg2000website,jpeg2000book}. The
incoherent remainder of both the DWPT and DWT converge rapidly
towards Gaussian statistics, while incoherent remainders of the
Fourier and JPEG do not converge to any value.

The coherent components of the DWPT and DWT retain all of the
properties of the total field, including the large-scale structures,
shape of the vorticity PDF, long spatial and temporal correlations,
and transport properties. Further, the coherent component contains
the large skewness and kurtosis of the PDF which are due to the
coherent structures. In contrast, the incoherent remainder has only
small-scale short-lived features and does not contribute
significantly to the transport. Thus in analysis of flow dynamics
and transport, it may be sufficient only to consider the coherent
component. These results suggest that it is reasonable to reduce the
computational complexity of turbulent flows by considering only the
low-dimensional coherent structures, which interact with a
statistically modelled incoherent background.

\begin{acknowledgements}~We thank Bruno Forisser and Emilie Regul for assistance in the
initial implementation of the Okubo-Weiss criterion and wavelet
analysis respectively. We thank Kenneth Ball, Carsten Beta, Ingrid
Daubechies, Marie Farge, Philip Marcus, Kai Schneider, and Jeffrey
Weiss for helpful discussions. This research was supported by the
Office of Naval Research.\end{acknowledgements}

\bibliographystyle{apsrev}

\end{document}